\documentclass[aps,pra, twocolumn, notitlepage, a4paper, superscriptaddress]{revtex4}
\usepackage[utf8]{inputenc}
\usepackage{amsmath, amssymb}
\usepackage{xcolor}
\usepackage{bm}
\usepackage{graphicx}

\begin{document}
\title{Quantum Kernel Methods for Solving Differential Equations}

\author{Annie E. Paine}
\affiliation{Pasqal SAS, 2 av. Augustin Fresnel, 91120 Palaiseau, France}
\affiliation{Department of Physics and Astronomy, University of Exeter, Stocker Road, Exeter EX4 4QL, United Kingdom}
\author{Vincent E. Elfving}
\affiliation{Pasqal SAS, 2 av. Augustin Fresnel, 91120 Palaiseau, France}

\author{Oleksandr Kyriienko}
\affiliation{Pasqal SAS, 2 av. Augustin Fresnel, 91120 Palaiseau, France}
\affiliation{Department of Physics and Astronomy, University of Exeter, Stocker Road, Exeter EX4 4QL, United Kingdom}

\date{\today}

\begin{abstract}
We propose several approaches for solving differential equations (DEs) with quantum kernel methods. We compose quantum models as weighted sums of kernel functions, where variables are encoded using feature maps and model derivatives are represented using automatic differentiation of quantum circuits. While previously quantum kernel methods primarily targeted classification tasks, here we consider their applicability to regression tasks, based on available data and differential constraints. We use two strategies to approach these problems. First, we devise a mixed model regression with a trial solution represented by kernel-based functions, which is trained to minimize a loss for specific differential constraints or datasets. Second, we use support vector regression that accounts for the structure of differential equations. The developed methods are capable of solving both linear and nonlinear systems. Contrary to prevailing hybrid variational approaches for parametrized quantum circuits, we perform training of the weights of the model classically. Under certain conditions this corresponds to a convex optimization problem, which can be solved with provable convergence to global optimum of the model. The proposed approaches also favor hardware implementations, as optimization only uses evaluated Gram matrices, but require quadratic number of function evaluations. We highlight trade-offs when comparing our methods to those based on variational quantum circuits such as the recently proposed differentiable quantum circuits (DQC) approach. The proposed methods offer potential quantum enhancement through the rich kernel representations using the power of quantum feature maps, and start the quest towards provably trainable quantum DE solvers. 
\end{abstract}
\maketitle

\section{Introduction}

Solvers of differential equations (DEs) are essential for all areas of science \cite{simmons2016differential, zachmanoglou1986introduction}. These include fluid dynamics, ecology, finance, medical science, and many more. While some simple instances of differential equations can be solved analytically, in majority of cases numerical solvers are required. Existing numerical methods heavily rely on finite differencing methods on finely discretized grids \cite{smith1985numerical}. Other classical methods include global spectral methods which effectively fit a function basis set to the differential equation problem considered \cite{boyd2001chebyshev}. Classical numerical solvers often suffer from instabilities that emerge in highly nonlinear systems. Another problem is the curse of dimensionality caused by an increase of grid for multidimensional systems. Thus, developing new techniques to solve DEs remains a hot area of contemporary research \cite{rackauckas2017differentialequations}, and increasingly requires new computational architectures \cite{rackauckas2019diffeqflux,Cai2022}.

Quantum computing offers advantages in performing certain computational tasks \cite{shor1994algorithms, harrow2009quantum, Arute2019}. Enabled by quantum principles, the use of superposition and entanglement can lead to fundamentally different scaling for quantum algorithms as compared to classical approaches \cite{NielsenChuang}. One example is for solving linear systems of equations \cite{harrow2009quantum}, and quantum speed-up of matrix-vector multiplication \cite{Biamonte2017}. When using finite-differencing, this translates directly to associated systems of DEs \cite{leyton2008quantum, Berry2017, lloyd2020quantum, LiuChilds2021, Jin2022, Linden2020}. However, previously described techniques rely on large-scale implementation of quantum phase estimation, and require resource overheads that render their implementation infeasible in foreseeable future \cite{Scherer2017}. This prompts to develop different approaches that can potentially help solving nonlinear DEs with near-term devices.

Recently, rapid improvement of quantum computing hardware has called for algorithms that can operate in the noisy regime \cite{Bharti2022}. In this case the hybrid quantum-classical workflow is often used. One possibility is to formulate a problem such that solution can be searched variationally \cite{cerezo2021variational}, where parameterized quantum circuits play a role similar to deep neural networks in classical machine learning (ML). This approach was coined as a quantum machine learning (QML) \cite{Benedetti2019,PerdomoOrtiz2018,Schuld2019QML}, and has triggered the development of various ML protocols for quantum hardware \cite{mitarai2018quantum,Liu2018,Zoufal2019,Coyle2020,Abbas2021,Du2021,Huang2021PRAppl,SChen2020a,Wu2021,SChen2021}. Variational approaches were also used for describing quantum evolution \cite{Endo2020,Cirstoiu2020} and linear algebra \cite{XU20212181,Bravo-Prieto2019,Chen2019}. In the field of nonlinear differential equations variational algorithms were used together with amplitude encoding \cite{lubasch2020variational}, where multiple quantum registers are required for encoding nonlinearity. Another approach was proposed in \cite{kyriienko2021solving}, where QML-type workflow is used. There a DE solution is represented by a differentiable quantum circuit (DQC), with nonlinear dependence being introduced via feature map encoding \cite{SchuldSweke2021} and cost function based readout, while function derivatives are introduced with the automatic circuit differentiation \cite{schuld2019evaluating,mitarai2018quantum}. Similar solutions were developed for continuous variable QML \cite{Knudsen2020}, stochastic differential equations \cite{Paine2021}, and generative modelling \cite{Romero2021,Kyriienko2022}.

Another facet of quantum machine learning was revealed when formulating models in terms of kernels---similarity functions that define a distance between two data points \cite{Havlicek2019,Schuld2019QML}. The core concept of kernel methods is the so-called `kernel trick' that maps data into a high-dimensional space \cite{ardeshir2021support}. Kernel methods are frequently used in classical machine learning, and aim to rewrite the ML task as a convex optimization problem \cite{kung2014kernel}. In the quantum domain, kernels are conveniently defined as overlaps between parametrized quantum states that represent data \cite{Havlicek2019,Schuld2019QML}, or any similar measure \cite{Huang2021}. It was conjectured that many supervised QML models can be considered as kernel methods that are well-suited to near term devices \cite{schuld2021quantum}. Currently these methods have mainly been considered for classification purposes \cite{mengoni2019kernel,li2015experimental}.

In this paper, we propose to use quantum kernel methods for regression problems, including solvers of nonlinear differential equations \cite{qk_patent}. In classical ML kernel methods are used for support vector regression (SVR) \cite{wang2005support}, where the kernel trick and convex optimization lead to expressive and provably trainable models. Kernel methods can also be applied to solve differential equations \cite{mehrkanoon2012approximate, mehrkanoon2015learning, lu2020solving}, while being limited by the expressivity of classical kernels. We describe two approaches that express solutions of DEs as quantum kernel-based models, and describe the rules for their automatic differentiation. We refer to the two as mixed model regression (MMR) and support vector regression (SVR) protocols. The protocols are applied to test problems of regression on quantum data, linear DEs, and nonlinear DEs in the form of Duffing equation \cite{Thompson_Stewart_2002}. We discuss the cases where the proposed workflow for DEs may provide advantage over existing methods.


\section{Quantum Kernel Methods}

We start by introducing the concept of a quantum kernel function. A kernel function is a conjugate-symmetric positive definite function $\kappa$ mapping two variables $x, y \in \mathcal{X}$ to the complex space, $\kappa : \mathcal{X} \times \mathcal{X} \rightarrow \mathbb{C}$. Quantum kernel function refers to a function which fulfils the requirements of a kernel function and can be evaluated by a quantum computer. 
An important result concerning kernel functions is known as the kernel trick. The kernel trick relies on the fact that any kernel function can be written as an inner product in a potentially high dimensional feature space, $\kappa (x,y) = \bm{\varphi}^\dag(x) \bm{\varphi}(y)$. Conversely $\bm{\varphi}^\dag(x) \bm{\varphi}(y)$ always represents a valid kernel function. This a consequence of Mercers theorem. Informally, it corresponds to the statement that for any symmetric positive definite function $f(s,t)$ there exists a countable set of functions $\{\phi_i\}_i$ such that $f(s,t)$ can be expressed as $f(s,t) = \sum_i \phi_i(s) \phi_i(t)$ \cite{mercer1909xvi}.
An example of a quantum kernel function is an overlap $\kappa (x,y) =  \langle \psi(x) | \psi(y) \rangle $ where $|\psi(x)\rangle$ denotes a state encoded by the variable $x$. This is an inner product which fulfils the requirements of a kernel function. Later we will consider other forms of the quantum kernel function, showing how to encode the variable into the state and how to evaluate the kernel. 

Our goal is to use the quantum kernel functions to solve differential equations. We consider two main methods --- mixed model regression (MMR) and support vector regression (SVR). Let us first consider using these methods to solve data-driven regression problems. This is a simpler case than solving differential equations yet still requires representing a solution function via quantum kernel function, and can be built upon to solve differential equations. For this regression problem we have a set of values $\{x_i, f_i\}_i$ and we want to find a function $f(x)$ that fits these points such that $f_i = f(x_i)$. We consider how both MMR and SVR approach the described problem. 


\subsection{MMR}

When using the mixed model regression we represent a trial function as
\begin{align}
\label{eq:MMRfunc}
    f_\alpha(x) = b + \sum_{i = 1}^{|\mathbf{y}|} \alpha_i \kappa(x, y_i),
\end{align}
where $\mathbf{y} = \{y_i\}_i$ is a set of evaluation points, and $\bm{\alpha} = \{\alpha_i\}_i$ and $b$ are tunable coefficients. We then write the problem defined by a loss function $\mathcal{L}(\alpha) = \sum_{i=1}^{|\mathbf{x}|} \left( f_\alpha(x_i) - f_i \right)^2$. This loss function is chosen such that when optimized with respect to $\alpha$ and $b$ the corresponding $f_\alpha$ solves the problem. 
The loss function requires the evaluation of $\{f_\alpha(x_i)\}_i$, which in turn requires the evaluation of $\{\kappa(x_i, y_j)\}_{i,j}$. These evaluations are independent of $\bm{\alpha}$ --- the variable which is adjusted during optimization. This means that the kernel function will only need to be evaluated once for each point in $\{x_i, y_j\}_{i,j}$ at the start of the optimization procedure. 
Any suitable optimisation method may be used to optimise $\mathcal{L}(\bm{\alpha})$. We can also see that the considered loss function is convex. 

We consider the general case $\mathcal{L}(\bm{\alpha}) = \sum_{i=1}^{|\mathbf{x}|} L(x_i ; \bm{\alpha})^2$ with $L$ being a linear function of $\bm{\alpha}$, and represents a distance. A sufficient condition for convexity of the loss function is $\partial^2 \mathcal{L} / \partial \alpha_j^2 \geq 0 $ everywhere for all $\alpha_j \in \bm{\alpha}$. We can write the second-order derivatives as
\begin{align}
\label{eq:lossconvexmid}
    \frac{\partial^2 \mathcal{L}}{\partial \alpha_j^2} &= \sum_{i=1}^{|\mathbf{x}|} \left[ 2 \left( \frac{\partial L}{\partial \alpha_j} \right)^2 + L  \frac{\partial^2 L}{\partial \alpha_j^2} \right] \\
    &= \sum_{i=1}^{|\mathbf{x}|} 2 \left( \frac{\partial L}{\partial \alpha_j} \right)^2  \geq 0 ,
\label{eq:lossconvexend}
\end{align}
where passing from Eq.~\eqref{eq:lossconvexmid} to Eq.~\eqref{eq:lossconvexend} we use the linearity of $L$ in $\alpha$.  When a loss function is convex its minimum is global, and there are bounds on convergence for various optimization methods \cite{boyd2004convex}. 
 
The workflow to solve an MMR problem is as follows:\vspace{1mm} \\
1. Choose setup for training, including the kernel function, optimizer, $\mathbf{x}$, $\mathbf{y}$.\\
2. Identify the loss function for problem considered.\\
3. Calculate set of kernel function evaluated over $\mathbf{x}  \otimes \mathbf{y}$.\\
4. Optimize the loss function.\vspace{1mm}

Once the model is trained, we can also evaluate it at a grid of points different from the training grid, learning the solution in the full domain of $x$.


\subsection{SVR}

For support vector regression we represent a trial function as $f(x) = \mathbf{w}^\dag \bm{\varphi}(x) + b$, where $\mathbf{w}$ and $b$ are tunable parameters, and $\bm{\varphi}(x)$ is a set of functions we later use the kernel trick upon. The first step is to write the problem as a primal (original) optimization model. This reads
\begin{align}
    &\mathrm{min}_{w, b, e} \{\mathbf{w}^T\mathbf{w} + \gamma \mathbf{e}^T \mathbf{e}\}, \\
    &\mathrm{subject~to}~f_i = \mathbf{w}^T \bm{\varphi}(x_i) + b + e_i, \quad i = 1:N,
\end{align}
where $\mathbf{e}$ is the set of error variables which relax the constraints from $f_i = \mathbf{w}^T \bm{\varphi}(x_i) + b + e_i$, and $\gamma$ is a tunable hyperparameter that changes the emphasis on minimising the error.

The process that follows is to write the model in its Lagrangian form, introducing a set of variables (known as dual variables) to implement each constraint. The Karush-Kuhn-Tucker (KKT) optimality conditions are then found, which emerge from equating the first derivative of the Lagrangian with respect to each of the primal and dual variables to zero \cite{kuhn2014nonlinear}. These conditions are then used to eliminate a subset of the primal variables. This can intuitively be understood as turning the variable into a constraint. This leads to a system of equations which have terms of $\mathbf{\varphi}(x_i)^T\mathbf{\varphi}(x_j)$, and by using the kernel trick these terms can be changed to $\kappa(x_i, x_j)$. Now the problem is written in a dual form as a system of equations to solve with coefficients involving kernel evaluations. Similar to the MMR method these have to be evaluated once at the start. The resulting system of equations is
\begin{align}
    \left[\begin{array}{c|c} \Omega +\hat{I}/\gamma & \mathbf{1} \\ \hline
    \mathbf{1}^T & 0 \end{array}\right]
    \left[\begin{array}{c} \bm{\alpha} \\ \hline
    b \end{array}\right] = 
    \left[\begin{array}{c} \mathbf{f} \\ \hline
    0 \end{array}\right],
    \label{eq:regSVR}
\end{align}
where $\Omega_{i,j} = \kappa(x_i, x_j)$, $\Omega = \{ \Omega_{i,j} \}_{i,j}$, and $\bm{\alpha}$ are a set of introduced dual variables. The system of equations can now be solved with any available method to solve such a problem. Once solved the relevant KKT conditions can be substituted into the expression $f(x; \alpha) = \mathbf{w}^\dag \mathbf{\varphi}(x) + b$, and the kernel trick applied to get an expression for $f(x)$ in terms of the dual variables, which have been solved for and kernel evaluations. We thus write our model as
\begin{align}
\label{eq:SVRfunc}
f(x; \alpha) = \sum_{i=1}^{|\mathbf{x}|} \alpha_i \kappa(x, x_i) + b.
\end{align}
Although we started considering $\bm{\varphi}$ as our fitting functions, the resulting function to this process is based on kernel evaluations and we never need knowledge of what $\bm{\varphi}$ are or to directly evaluate them.

The workflow to prepare an SVR problem is as follows:\vspace{1mm} \\
1. Write model with minimization function and constraints.\\
2. Write out Lagrangian.\\
3. Find the KKT optimality conditions.\\
4. Eliminate subset of original optimization variables.\\
5. Use the kernel trick to realise problem in terms of kernels.\\
6. Write out remaining relationships as system of equations.\\
7. Use KKT conditions and kernel trick to express function in terms of kernel functions.\vspace{1mm}

In the Appendix this process is worked through in more detail for a specific (DE) example. The prepared SVR model can then be used for any problem of the form assumed in preparing the original model. The workflow for solving an SVR problem is as follows:\vspace{1mm} \\
1. Choose setup for training, including the kernel function, system of equations solver, $\mathbf{x}$, $\gamma$.\\
2. Identify suitable SVR model for the problem considered.\\
3. Calculate set of kernel function evaluated over $\mathbf{x}  \otimes \mathbf{x}$.\\
4. Solve system of equations.\vspace{1mm}

We also note that SVR method results in a form that can still be considered as an optimization problem to be solved with an optimizer. The system of equations $Ax = b$ can be translated into the loss function $\mathcal{L}(x) = \sum_i [(Ax)_i - b_i]^2$. Here we use MSE loss but other forms can be employed. This formulation can especially useful when considering problems resulting in nonlinear systems of equations.

Comparing the MMR and the SVR methods we note that the solving workflow for the two are similar. Namely, we choose a setup, identify what to solve based on method and problem, calculate the set of kernel function evaluations, and solve the model identified in step two. However, identifying the model for the SVR method is a more involved process. 

Both MMR and SVR result in a function approximation to the solution of the problem considered. For regression this is Eq.~\eqref{eq:MMRfunc} and Eq.~\eqref{eq:SVRfunc}, respectively. These two functions look very similar with the difference being the kernel evaluation at $y_i$ for MMR versus $x_i$ for SVR. This is a consequence of using the kernel trick when formulating the SVR model, which necessarily results in $\mathbf{y} = \mathbf{x}$. Also to be highlighted is that the form of Eq.~\eqref{eq:SVRfunc} depends on the problem considered. For example, later we see that when solving differential equations, evaluations of the kernel derivative are involved in the function expression. However, for MMR the form of the model remains the same no matter what problem considered.

One benefit of using the MMR model is the simpler identifying of the model to solve. Another is the convexity when considering certain problems. The benefit of SVR is that when linear the resulting system of equations to solve is also linear and thus has deterministic solution. Furthermore, the initial trial function is in terms of $\varphi$ which can be of higher dimensionality than the kernel function yet never needs to be evaluated directly.


\section{Quantum Kernel Function Evaluation}

We now look further into the specifics of quantum kernel functions. In particular, we consider their structure, where feature maps encode dependence on a variable into a state. We also consider how to evaluate them and their derivatives. Earlier we mentioned a quantum kernel function of the form $\kappa (x,y) =  \langle \psi(x) | \psi(y) \rangle$, being an inner product that is generally complex for quantum states. In the following, we consider $\kappa (x,y) =  |\langle \psi(x) | \psi(y) \rangle|^2$ as an absolute value square of the overlap. This also corresponds to a valid kernel function \cite{schuld2021quantum}. We consider this kernel function as it is real valued --- an advantage when expressing real valued functions.
\begin{figure}
\begin{center}
\includegraphics[width=1.0\linewidth]{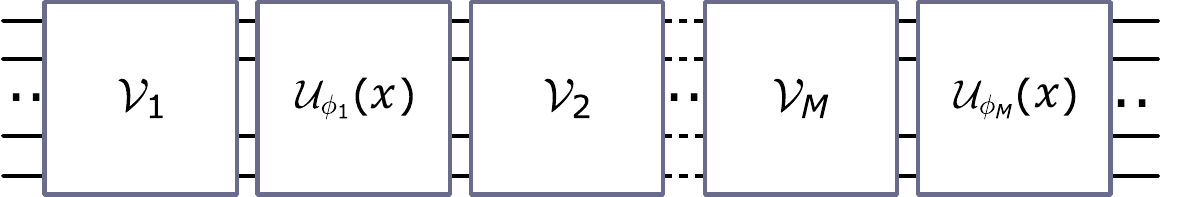}
\end{center}
\caption{Circuit diagram showing a general form of function encoding circuit $\mathcal{U}(x)$ used to implement quantum kernel. This is formed by layers of static circuits $\mathcal{V}_i$ and data re-uploading circuits $\mathcal{U}_{\phi_i}(x)$ parametrized by $x$.}
\label{fig:enc_circ}
\end{figure}

\subsection{Encoding}

The kernel functions we consider contain states $|x\rangle$, which are encoded by a classical variable $x$. To create such states we use feature map encoding, where $x$ is embedded into the state by parametrizing gates preparing the state, $|\psi(x)\rangle = \mathcal{U}(x)|0\rangle$. A simple example of $\mathcal{U}(x)$ is the product feature map $\mathcal{U}_{\phi}(x) = \bigotimes_{j=1}^{N} R_{\alpha,j} (\phi(x))$, where $R_{\alpha,j}(\phi(x))$ is rotation on qubit $j$ of angle $\phi(x)$ about a Pauli operator $\alpha$. Other more complicated feature map encodings can be considered. The generalization may include the re-uploading technique \cite{PerezSalinas2020datareuploading} where action of feature maps can be layered with (non-variational) entangling circuits, $\mathcal{U}(x) = \mathcal{U}_{\phi_M}(x) \mathcal{V}_M ... \mathcal{V}_2 \mathcal{U}_{\phi_1}(x) \mathcal{V}_1$ (see Fig. \ref{fig:enc_circ}). This layered form terminates with a circuit encoded by a variable as a final entangling circuit cancels out for kernels based on $\mathcal{U}^\dag(x)\mathcal{U}(y)$. As with many variational algorithms, when choosing feature maps it is important to have a map expressible enough to represent the solution to the problem whilst also being trainable \cite{Caro2021encodingdependent}.


\subsection{Evaluation}

\begin{figure}
\begin{center}
\includegraphics[width=1.0\linewidth]{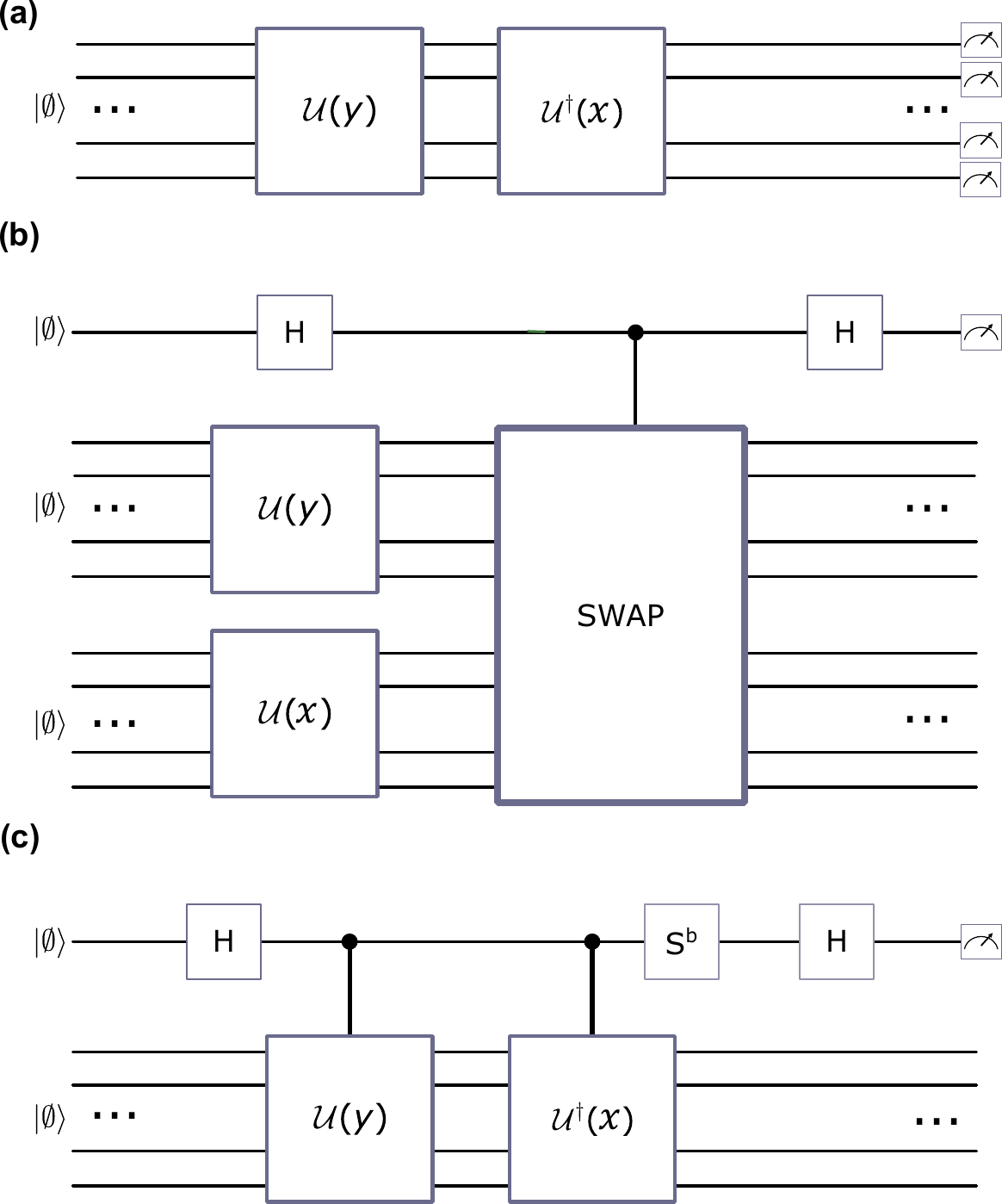}
\end{center}
\caption{Circuit diagrams for evaluating the kernel $\kappa(x,y) = |\langle \psi(x) | \psi(y) \rangle|^2$, where in all circuits $\mathcal{U}$ and $H$ represent the kernel feature map and the Hadamard gate, respectively. (a) Naive kernel evaluation based on consecutive application of $\mathcal{U}$ circuits, followed by measuring each qubit. The kernel value is inferred from a probability of returning to the initial state. (b) SWAP test measuring $|\langle \psi(x) | \psi(y) \rangle|^2$. The controlled SWAP onto the size $2N$ register is composed of qubitwise controlled SWAP on the $n^{th}$ qubit pair, repeated for $n \in 1:N$. (c) Hadamard test measuring Re($\langle 0 |\mathcal{U}^\dag (x) \mathcal{U}(y) | 0 \rangle$) and Im($\langle 0 |\mathcal{U}^\dag (x) \mathcal{U}(y) | 0 \rangle$) for $b=0$ and $b=1$, respectively. $\mathrm{S}$ denotes the phase gate, $\mathrm{exp}(-\pi Z/4)$. }
\label{fig:Keval}
\end{figure}

We now discuss how to implement the quantum kernel function $\kappa (x,y) =  |\langle \psi(x) | \psi(y) \rangle|^2 $. One way is to use the coherent SWAP test \cite{buhrman2001quantum,Higgott2019variationalquantum}. This test requires $2N+1$ qubits, where $N$ is the number of qubits used to express $|\psi(x) \rangle$. $|\psi(x)\rangle$ and $|\psi(y)\rangle$ are both prepared on separate registers then via Hadamard gates and controlled operations an ancillary qubit can then be measured to read $|\langle \psi(x) | \psi(y) \rangle|^2$. The circuit diagram is shown in Fig.~\ref{fig:Keval}(b).

We can also employ other methods. For this, we use the fact that the kernel evaluation can be written as
\begin{align}
    |\langle \psi(x) | \psi(y) \rangle|^2 &=  \langle 0| \mathcal{U}^\dag (y) \mathcal{U}(x) |0 \rangle \langle 0 |\mathcal{U}^\dag (x) \mathcal{U}(y) | 0 \rangle. \label{eq:kernelexp}
\end{align}
The measurement in Eq.~\eqref{eq:kernelexp} can be implemented naively by the circuit in Fig.~\ref{fig:Keval}(a). The circuit is initialized in the zero state. Then $\mathcal{U}(y)$ is applied, followed by $\mathcal{U}^\dag(x)$. The probability of remaining in the zero state and thus the kernel is then calculated my measuring all qubits and finding the ratio of times $|0\rangle$ is measured.

Another possible implementation is two evaluations of the Hadamard test with $N+1$ qubits as shown in Fig.~\ref{fig:Keval}(c) \cite{Mitarai2019Htest}. This can be used to evaluate the real and imaginary parts of $\langle 0 |\mathcal{U}^\dag (x) \mathcal{U}(y) | 0 \rangle$ which can then be used to evaluate the kernel as $\mathrm{Re}(\langle 0 |\mathcal{U}^\dag (x) \mathcal{U}(y) | 0 \rangle)^2 + \mathrm{Im}(\langle 0 |\mathcal{U}^\dag (x) \mathcal{U}(y) | 0 \rangle)^2$.

\subsection{Derivatives}
As our goal is to solve differential equations, we need to be able to evaluate derivatives of the kernel function. We introduce notation for the derivatives as follows,
\begin{align}
    \nabla^m_n \kappa(x,y) = \frac{\partial^{m+n} \kappa(x,y)}{\partial x^n \partial y^m}.
\end{align}

\begin{figure}
\begin{center}
\includegraphics[width=1.0\linewidth]{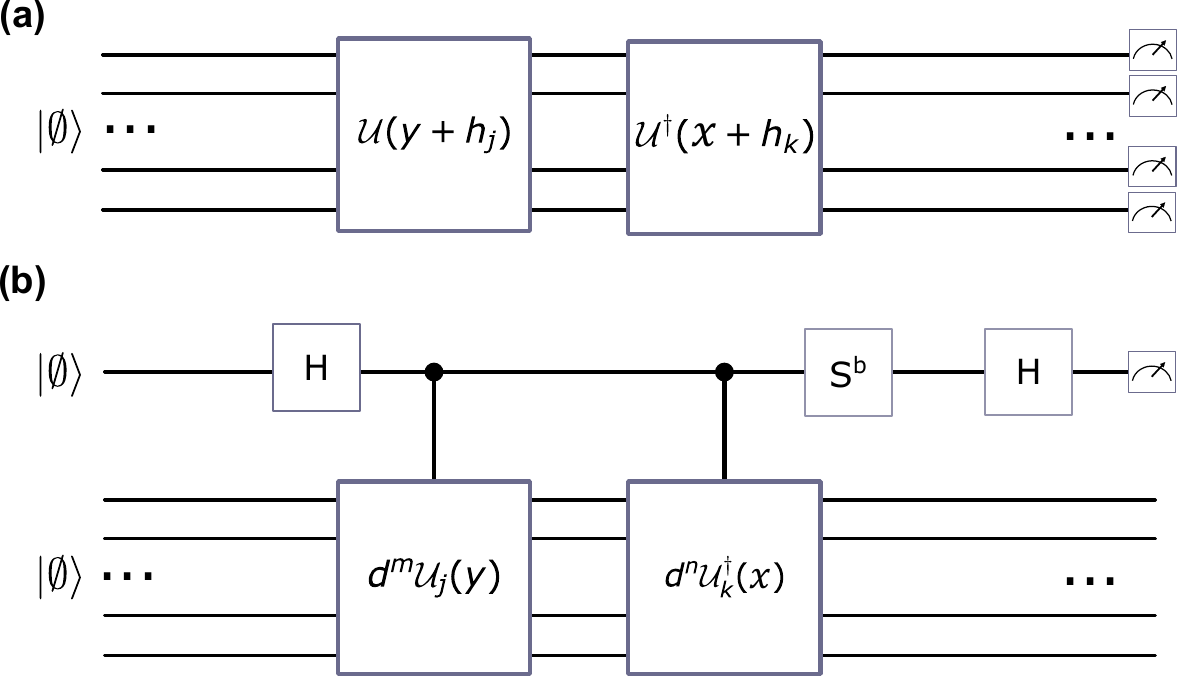}
\end{center}
\caption{Circuit diagrams used for evaluation of derivatives. (a) Generic circuit for differentiating kernels shown in Fig.~\ref{fig:Keval}(a) where a parameter shift rule is used. Depending on which derivative is calculated, gates parametrized by $x$ and/or $y$ have their parameters shifted up and down. Contributions from all parametrized gates are then summed for overall derivative. (b) Using Hadamard test for evaluation of the overlap $\langle 0 | d^n/dx^n \mathcal{U}_k^\dagger(x) d^m/dy^m \mathcal{U}_j(y) | 0 \rangle$, where $k$ and $j$ index over which gates with $x$ and/or $y$ as parameters are differentiated. When $b=0$ and $b=1$ are used the real and imaginary part is evaluated. By summing over $j,k$ the full overlap $\langle 0 | d^n/dx^n \mathcal{U}^\dagger(x) d^m/dy^m \mathcal{U}(y) | 0 \rangle$ can be evaluated. These overlaps can then be used to evaluate kernel derivatives.}
\label{fig:Deval}
\end{figure}

To implement derivative evaluation, one way is to consider the kernel as written in Eq.~\eqref{eq:kernelexp} and the parameter shift rule \cite{schuld2019evaluating,mitarai2018quantum}. With this method we take the kernel evaluation method as in Fig.~\ref{fig:Keval}(a) but shift $x$ and $y$ up and down depending on what derivative is being calculated in each gate that they parametrize. For example for the first order derivative with respect to $x$ the number of evaluations of Fig.~\ref{fig:Deval}(a) is $2n$ with $n$ being the number of gates parametrized by $x$. Using the parameter shift rule means we calculate the analytic derivative though it does place some requirements on the gates parametrized by $x/y$ such as being involutory. Generalized parameter shift rules are possible, where such requirements are relaxed \cite{kyriienko2021generalized, wierichs2021general,Izmaylov2021generalized,Vidal2018,Theis2021}.

We can also implement derivatives via the Hadamard test. First, we note the form of the first-order derivative of the kernel in $x$ by using the product rule in Eq.~\eqref{eq:kernelexp} as 
\begin{align}
\notag
    &\frac{\partial}{\partial x} \kappa(x,y) = \langle 0| \mathcal{U}^\dag (y) d/dx(\mathcal{U}(x)) |0 \rangle \langle 0 |\mathcal{U}^\dag (x) \mathcal{U}(y) | 0 \rangle \\ &+ \langle 0| \mathcal{U}^\dag (y) \mathcal{U}(x) |0 \rangle \langle 0 |d/dx(\mathcal{U}^\dag (x)) \mathcal{U}(y) | 0 \rangle,
\end{align}
and thus we can evaluate this derivative by evaluating $\langle 0| \mathcal{U}^\dag (y) d/dx(\mathcal{U}(x)) |0 \rangle$ and  $\langle 0 |\mathcal{U}^\dag (x) \mathcal{U}(y) | 0 \rangle$ (real and imaginary parts). The second term can be evaluated as for evaluating the kernel shown in Fig.~\ref{fig:Keval}(b), and calculations can be reused because the derivatives are evaluated over the same set of points as the kernel itself. To calculate the first term a modified Hadamard test can be used. 

We consider the generalized layered form of kernel encoding $\mathcal{U}(x) = \mathcal{U}_{\phi_M}(x) \mathcal{V}_M ... \mathcal{V}_2 \mathcal{U}_{\phi_1}(x) \mathcal{V}_1$ with each feature map being $U_{\phi_j}(x) = \mathrm{exp}(-i\mathcal{G}_j \phi_j(x))$. For this case the derivative reads $U'(x) = \sum_{j=1}^M \mathcal{U}_{M:j+1} \mathcal{U}_{\phi_j}(-i\mathcal{G}_j) \phi_j'(x) \mathcal{V}_j \mathcal{U}_{j-1:1}$ with $\mathcal{U}_{j:k} = \mathcal{U}_{\phi_{j}}(x) \mathcal{V}_{j} \mathcal{U}_{\phi_{j-1}}(x) ... \mathcal{U}_{\phi_k}(x)\mathcal{V}_k$. We can now assume that the generators $\mathcal{G}_j$ are unitary. When $\mathcal{G}_j$ are unitary we can calculate each overlap term in the derivative expansion with two Hadamard tests. However, if this is not the case, one can decompose them into sums of unitary terms and evaluate them separately with increased number of Hadamard tests \cite{schuld2019evaluating}. 

Once the procedure for evaluating derivatives being set up, we generalize to higher-order derivatives. By using the product rule in Eq.~\eqref{eq:kernelexp} whatever derivative is required, one can express it as sums of products of overlaps with $\mathcal{U}(x)$ and $\mathcal{U}(y)$ differentiated to different orders. These overlaps can be calculated with two (when generators unitary) overlap tests for each gate with $x$ and/or $y$ as a parameter (see Fig.~\ref{fig:Deval}). These overlap evaluations can be reused for calculating different derivatives where the same overlap occurs.


\section{Solving Differential Equations}

In the following, we collect the described tools for model and derivative evaluations, and apply them to solve differential equations. While there are many possible choices, we start by considering a simple class given by the differential constraint
\begin{equation}\label{eq:simple_de}
\mathrm{DE}(x, f, df/dx) = \frac{df}{dx} - g(x, f) = 0,
\end{equation}
with initial condition $f(x_0) = f_0$, and $g$ a smooth function of $x$ and $f$ which in general can be nonlinear in either of those arguments. We now use both MMR and SVR to solve this type of DEs in next subsections.


\subsection{MMR}

When solving DEs of the type \eqref{eq:simple_de} via MMR, we choose a loss function in the form $\mathcal{L}(\alpha)= \sum_{i=1}^{|x|}\left[ \mathrm{DE}(x_i, f_\alpha(x_i), df_\alpha/dx(x_i)) \right]^2 + (f_\alpha(x_0) - f_0)^2$. We remind that the trial function reads $f_\alpha(x) = b + \sum_{i = 1}^{|y|} \alpha_i \kappa(x, y_i)$. Therefore, kernels $\kappa$ and their derivatives $\nabla_1^0 \kappa$ are evaluated over $\{x_i, y_j\}_{i,j}$, leading to corresponding $f_\alpha$ and $df_\alpha/dx$ evaluations. These values are independent of $\alpha$, and only need to be evaluated once at the start, then being reused throughout optimization. The loss function can the be optimized via any appropriate optimizer for getting optimal weights $\alpha_{\mathrm{opt}}$. The resulting function is then a suitable approximation to the solution of the differential equation, mainly being limited by expressivity of the model and generalization bounds. 

When the differential equation is linear [i.e., $g$ is linear in $f$ in Eq.~\eqref{eq:simple_de}] the considered loss function is convex. This is true when the differential equation is linear and $f_\alpha$ (and consequently $f_\alpha'$) is linear in $\alpha$, meaning we are in the situation as described by Eq.~\eqref{eq:lossconvexend}. When the differential equation is nonlinear this is \emph{not necessarily} the case. In order to determine that one needs to check for the convexity of the loss function. One possibility is a numerical check by sampling the second derivatives of the loss with respect to the optimizable parameters at many locations. If this value is ever negative then the problem is non-convex.


\subsection{SVR}

When considering solving DEs with support vector regression, the formulation of the problem changes depending on the form of differential equation considered \cite{mehrkanoon2012approximate, mehrkanoon2015learning, lu2020solving}. The steps for the problem formulation however remain the same: state a model, write out Lagrangian, find KKT optimality conditions, eliminate subset of prime variables by using the KKT conditions, use the kernel trick, and finally write out remaining equations in matrix form. 

We follow the SVR formulation procedure for problems of the form $\mathrm{DE}(x, f, df/dx) = df/dx - g(x, f) = 0$ with initial condition $f(x_0) = f_0$. We provide the details in the Appendix, and here provide the resulting set of equations in the matrix form:
\begin{align}
    \left[\begin{array}{c|c|c|c|c} \tilde{\Omega}_1^1 & \Omega_0^{1} & \mathbf{h}_0^1 & \mathbf{0} & \hat{0} \\
    \hline
    \Omega_1^{0} & \tilde{\Omega_0^0} & \mathbf{h}_0^0 & \mathbf{1} & -I \\
    \hline
    {\mathbf{h}^T}_1^0 & {\mathbf{h}^T}_0^{0} & \tilde{h} & 1 & \mathbf{0}^T \\
    \hline
    \mathbf{0}^T & \mathbf{1}^T & 1 & 0 & \mathbf{0}^T \\
    \hline
    \hat{D} & I & \mathbf{0} & \mathbf{0} & \mathbf{0}
    \end{array}\right] \left[ \begin{array}{c} 
    \bm{\alpha} \\ \hline \bm{\eta} \\ \hline \beta \\ \hline b \\ \hline \mathbf{y}
    \end{array} \right]= 
    \left[ \begin{array}{c} 
    \mathbf{\tilde{g}} \\ \hline \mathbf{0} \\ \hline f_0 \\ \hline 0 \\ \hline \hat{0}
    \end{array} \right].
\end{align}
Here, we introduced dummy variables $y_i$. $\bm{\eta}$ and $\beta$ are dual variables introduced along with $\bm{\alpha}$ corresponding to the dummy variable constraint and the initial variable constraint, respectively. The remaining notation is as follows
\begin{align}
    [\Omega^m_n]_{i,j} &= \nabla^m_n \kappa(x_j, x_i), \\
    \tilde{\Omega}^m_n &= \Omega^m_n + \hat{I}/\gamma , \\
    [\mathbf{h}^m_n]_i &= \nabla^m_n \kappa(x_0, x_i) , \\
    \tilde{h} &= \kappa(x_0, x_0), \\
    \hat{D} &= \mathrm{diag}\left( \left\{ \frac{\partial g}{\partial f}(x_i, y_i) \right \}_i \right), \\
    [\mathbf{\tilde{g}}]_i &= g(x_i, y_i) .
\end{align}

We now have a set of generally nonlinear equations, which can be solved for finding a vector of optimized weights. By substituting the relevant KKT optimality conditions into $f(x) = b + \sum_{i = 1}^{|y|} \alpha_i \kappa(x, y_i)$ and employing the kernel trick, we get an expression for $f$ in terms of kernel functions
\begin{align}
    f(x) = \sum_{i=1}^{|\mathbf{x}|} \alpha_i \nabla_1^0 \kappa(x_i, x) + \sum_{i=1}^{|\mathbf{x}|} \eta_i \kappa(x_i, x) + \beta \kappa(x_0, x) + b,
\end{align}
where optimized variables (weights) are used. Note that if the differential equation is linear, the dummy variable constraints of $y_i$ are not required. This leads to a system of linear equations with lower dimension. 


\subsection{Other forms of DEs}

Many practical problems are not of the form $\mathrm{DE}(x, f, df/dx) = df/dx - g(x, f) = 0$ considered above. For instance, they may include terms of higher order, higher dimension, or indeed many other different variations. When considering the MMR method, one can readily generalize is to any other form of DE simply relying on generalized optimization. For this, a suitable loss function needs to be formulated for the chosen equation. Additionally, we shall be able to evaluate each term of the differential equation. For systems of DEs the overall loss becomes the sum of the loss of each individual differential equation within the system. For PDEs with domains of more than one dimension, the kernel function can be considered as $\kappa(\mathbf{x}, \mathbf{y}) = |\langle 0 |\mathcal{U}^\dag (\mathbf{x}) \mathcal{U}(\mathbf{y}) | 0 \rangle|^2 $, where the feature maps now encode a vector of domain variables. The simplest form is $\mathcal{U} (\mathbf{x}) = \mathcal{U} (x_1) \mathcal{U} (x_2) ... \mathcal{U} (x_M)$ with $M = |\mathbf{x}|$. 

When the SVR method is used, the considered problem needs to be formulated into the SVR form, resulting in a different form of matrix equation. Higher order derivative SVRs \cite{mehrkanoon2012approximate, lu2020solving} and SVRs for PDEs \cite{mehrkanoon2015learning} are possible, as well as their generalizations for systems of differential equations.


\section{Results}

Having established quantum kernel approaches for solving DEs and learning from data, we apply them to specific problems and show the results. 

\textit{Regression on quantum data.} We start by considering the case of regression. We generate a quantum dataset that corresponds to dynamics of total magnetization of a biased honeycomb Kitaev model \cite{Savary2016,Hermanns2018}. The Hamiltonian of the system reads
\begin{align}
\notag
\mathcal{H} = &J (\sum_{\langle i,j\rangle \in \mathcal{X}} X_{i} X_{j} + \sum_{\langle i,j\rangle \in \mathcal{Y}} Y_{i} Y_{j} + \sum_{\langle i,j\rangle \in \mathcal{Z}} Z_{i} Z_{j}) \\ + & h_z \sum_{j=1}^{N} \hat{Z}_{j},
\label{eq:H_Kitaev}
\end{align}
where $\mathcal{X}$, $\mathcal{Y}$, $\mathcal{Z}$ are sets of bonds. We choose antiferromagnetic coupling and set $h_z/J = 0.2$. Specifically, we simulate nonequilibrium effects by performing time evolution of $M_z = \sum_j Z_j/N$ for $N=12$ qubits on a lattice with periodic boundary conditions \cite{bespalova2021quantum}, starting from the uniform initial state. The choice of quantum dataset with strong magnetic correlations may be especially suitable for kernel-based regression, given recent advances in learning from experiments \cite{Huang2021b}.
Choosing a subset of evolved magnetization values labeled by $x$ values (here corresponds to time), we proceed to perform MMR.
\begin{figure}
\begin{center}
\includegraphics[width=1.0\linewidth]{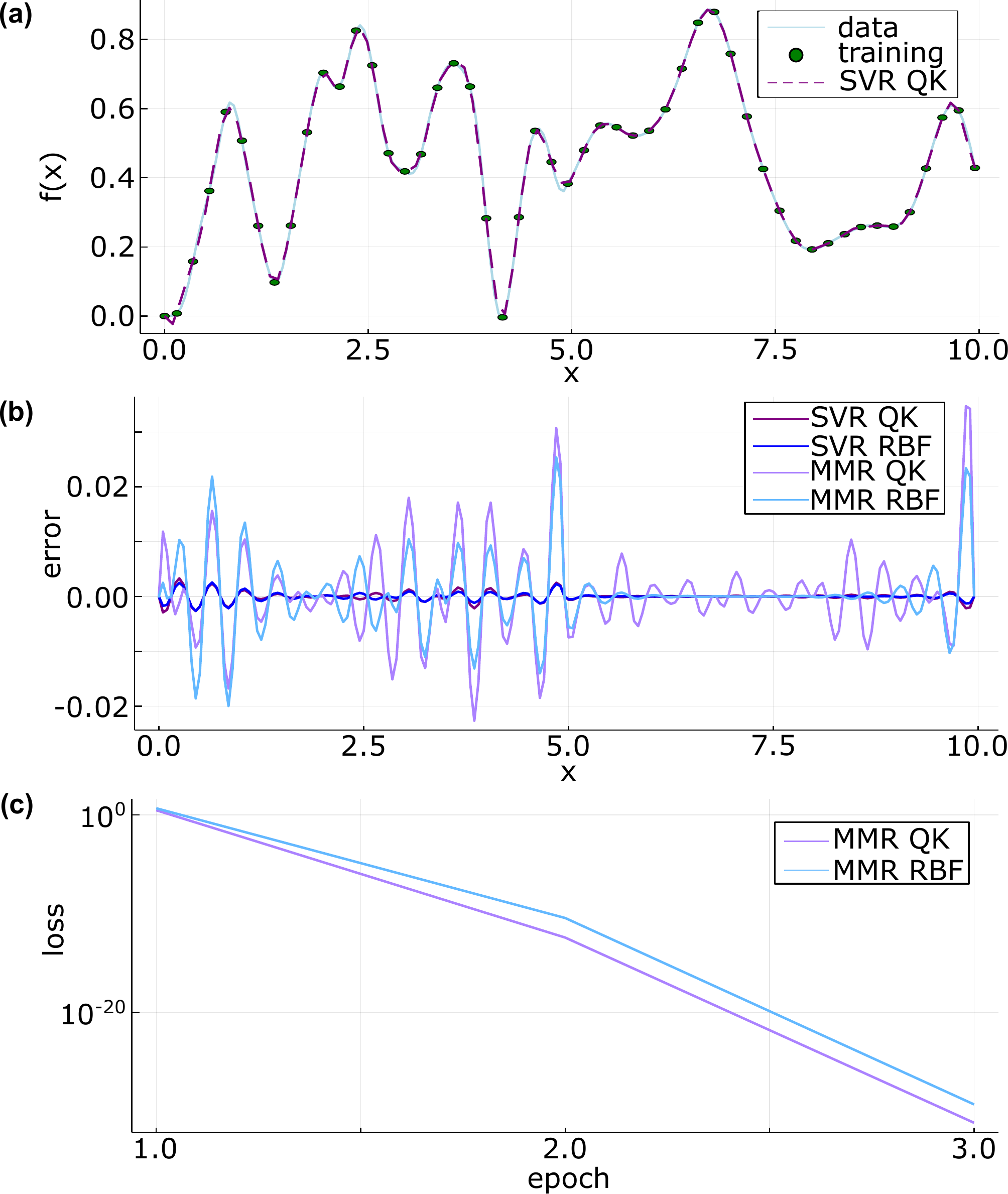}
\end{center}
\caption{MMR and SVR used to solve a regression problem. Data to fit is for time-evolved magnetization of a biased honeycomb Kitaev model. (a) Solution via SVR method for quantum kernel with two layers and $N=8$ shown by dashed purple line. Data plotted as solid light blue curve with points used for training highlighted with green circles. (b) Error between results and underlying data plotted over $x$ as $[f(x) - f_{\mathrm{data}}(x)]$, and we additionally normalize data by the range of magnetization values. Error plotted for result shown in (a) and SVR method with classical RBF kernel with $\sigma = 0.2$. Also plotted results from MMR method with same kernels considered. Newton optimizer with just $3$ epochs is used for MMR method. (c) Loss value over epoch number plotted for MMR results shown in (a) and (b).}
\label{fig:reg}
\end{figure}

When implementing the MMR method we consider $\mathbf{x}$ with $51$ values of $x$ between $1$ and $10$, associated to the data, and $\mathbf{y} = \mathbf{x}$. We use and compare the results from a classical kernel and a quantum kernel. The classical kernel used is a commonly used radial basis function (RBF) kernel $\kappa(x, y)=\exp\left[(x-y)^2/(2\sigma^2)\right]$, with $\sigma$ being a hyperparameter that describes a width of the kernel. In calculations we choose $\sigma =0.2$ as that shows favorable performance. For the quantum kernel, we use layers of depth-five HEA and feature maps based on parametrized X rotations, $R_x (\phi(x))$, acting on each qubit. We set $\phi(x) = q x / 2$, where $q$ is the qubit index, and consider a register of eight qubits. For the loss function MSE is used with a pinned boundary formulation (see \cite{kyriienko2021solving} for the details of boundary handling). The loss function for data regression is convex and is optimized via Newton's method. In this case just three epochs is enough for converging to low loss values. We model the system with full state simulation using the \texttt{Julia}'s package \texttt{Yao.jl} \cite{YaoFramework2019}. The error of the results of this are shown in Fig.~\ref{fig:reg}(b) with associated loss in Fig.~\ref{fig:reg}(a). As can be seen, both kernel types are able to closely approximate the considered function. Moreover, we note that for complicated quantum data coming from spin-spin correlation one can benefit from specifically-designed quantum kernels that account for the structure of the problem.

When implementing the SVR method, we use the same points $\mathbf{x}$, kernel functions and the simulation package. The resulting SVR system of equations to solve for this form of problem are as shown in Eq.~\eqref{eq:regSVR}. The results of this with the quantum kernel are shown in Fig.~\ref{fig:reg}(a). The error of the results is shown in Fig.~\ref{fig:reg}(b). It can be seen that both kernel types are able to closely approximate the considered function, and that SVR outperforms the MMR method.

\textit{Linear DEs.} Next, we consider solvers of linear differential equations. In particular, we solve the equation
\begin{align}
    \frac{df}{dx} = - \lambda \kappa f - \lambda \mathrm{exp}(- \lambda \kappa x ) \mathrm{sin}(\lambda x),
    \label{eq:linearde}
\end{align}
where parameters are chosen as $\lambda = 20$ and $\kappa = 0.1$, along with initial condition $f(0) = 1$. The analytic solution to the differential equation \eqref{eq:linearde} is $f_\mathrm{sol}(x) = \mathrm{exp}(- \lambda \kappa x ) \mathrm{cos}(\lambda x)$, being a fading oscillatory dependence.
\begin{figure}
\begin{center}
\includegraphics[width=1.0\linewidth]{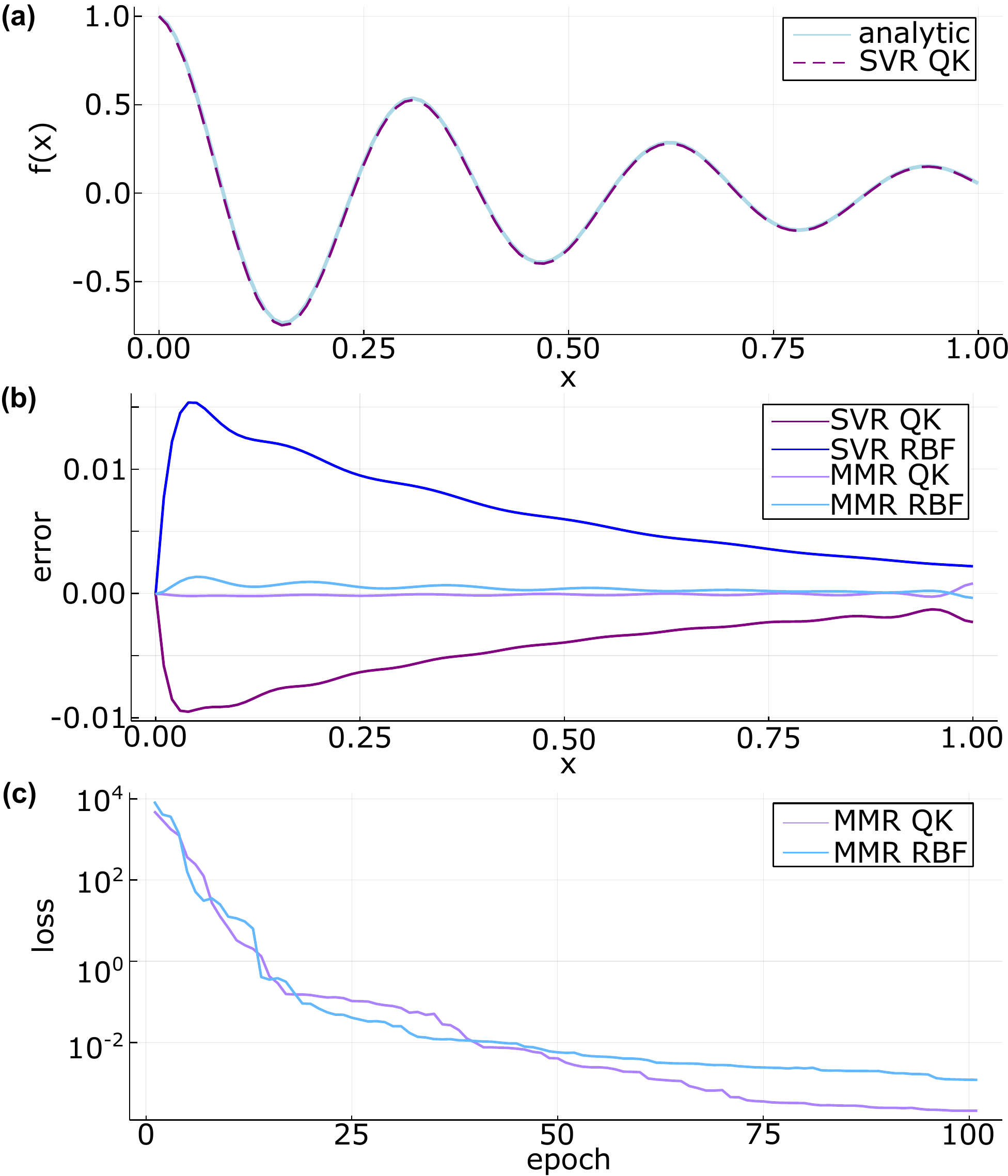}
\end{center}
\caption{MMR and SVR used to solve a linear differential equation \eqref{eq:linearde}. $\lambda = 20$, $\kappa = 0.1$ and $f(0) = 1$. (a) Solution via SVR method for quantum kernel with two layers and $N=8$ shown by dashed purple line. Known analytic solution plotted with solid light blue line. (b) Error between results and analytic solution plotted over x as $(f(x) - f_{\mathrm{sol}}(x))/\mathrm{range}(f_\mathrm{sol})$. Error plotted for result shown in (a) and SVR method with classical RBF kernel with $\sigma$ = 0.2. Also plotted results from MMR method with same kernels considered. Newton optimiser with $100$ epochs used for MMR method. (c) Loss value over epoch number plotted for MMR results shown in (a) and (b).}
\label{fig:linearde}
\end{figure}

When implementing the MMR method we consider $\mathbf{x}$ and $\mathbf{y}$ of $20$ points uniformly spaced over $[0,1]$. We use and compare the results from a classical RBF kernel with $\sigma = 0.2$ and a quantum kernel with two layers of HEA circuits (depth equal to five) followed by feature map of $R_x (\phi(x))$ on each qubit with $\phi(x) = q x / 2$, where $q$ is the qubit index. We consider eight qubits in the register. For the loss function MSE is used with a pinned boundary. This loss function is convex, as the DE is linear, and is optimized via Newtons method. The error of the results are shown in Fig.~\ref{fig:linearde}(b) with corresponding loss in Fig.~\ref{fig:linearde}(c). As can be seen both kernel types are able to closely approximate the considered function with the error less than $0.002$ in magnitude, the quantum kernel slightly outperforms the classical kernel although we did not further explore hyperparameter optimization.

When implementing the SVR method, we use the same $\mathbf{x}$ and kernel functions. The corresponding SVR system of equations to solve for a problem of form $df/dx + g(x)f + r(x) = 0$ reads
\begin{align}
    \left[\begin{array}{c|c|c} \hat{M}  & \mathbf{h}_1^0 - \hat{D} \mathbf{h}_0^0 & \mathbf{g}  \\
    \hline
    (\mathbf{h}_1^0 - \hat{D} \mathbf{h}_0^0)^T & \tilde{h}_0^0 & 1 \\
    \hline
    \mathbf{g}^T & 1 & 0  \\
    \end{array}\right]
    \left[ \begin{array}{c} 
    \bm{\alpha} \\ \hline \beta \\ \hline b 
    \end{array} \right]= 
    \left[ \begin{array}{c} 
    \mathbf{\tilde{r}} \\ \hline f_0 \\ \hline 0 
    \end{array} \right],
\end{align}
where the notation is as follows:
\begin{align}
    [\Omega^m_n]_{i,j} &= \nabla^m_n \kappa(x_j, x_i), \\
    \tilde{\Omega}^m_n &= \Omega^m_n + I/\gamma ,\\
    [\mathbf{h}^m_n]_i &= \nabla^m_n \kappa(x_0, x_i), \\
    \tilde{h}^m_n &= \kappa(x_0, x_0)^m_n ,\\
    \hat{D} &= \mathrm{diag}\left( \left\{ g(x_i) \right\}_i \right),\\
    [\mathbf{\tilde{g}}]_i &= g(x_i), \\
    [\mathbf{\tilde{r}}]_i &= r(x_i), \\
    \hat{M} &= \Omega_1^1 - \Omega_1^0 \hat{D} - \hat{D} \Omega_0^1 + \hat{D} \tilde{\Omega}_0^0 \hat{D}.
\end{align}
We choose $\gamma = 10^5$ and this system is then solved with \texttt{Julia}'s built-in matrix-defined linear equation solver. The error of these results is shown in Fig.~\ref{fig:linearde}(b), with the result from using the quantum kernel shown explicitly in Fig.~\ref{fig:linearde}(a). Again it can be seen that both kernel types are able to closely approximate the considered function, though not as closely as the MMR method, with quantum kernel outperforming the classical kernel.

\textit{Solving nonlinear DEs.} We now move on to consider solving \emph{nonlinear} differential equations. We show the results of solving the non-damped Duffing equation \cite{Thompson_Stewart_2002}
\begin{align}
    \frac{d^2f}{dx^2} = c ~  \mathrm{cos}(dx) - a f - b f^3.
    \label{eq:duffing}
\end{align}
We consider $a=1, b=1, c=3, d=3$ as well as initial conditions $f(0) = 1$ and $f'(0) = 1$, and solve with both MMR and SVR methods. To compare our solutions we also solve the problem with a classical numerical technique with the \texttt{Julia}'s package \texttt{DifferentialEquations.jl} \cite{rackauckas2017differentialequations} specifying to solve the problem with a fifth-order Tsitouras method [Tsit5(...)].

\begin{figure}
\begin{center}
\includegraphics[width=1.0\linewidth]{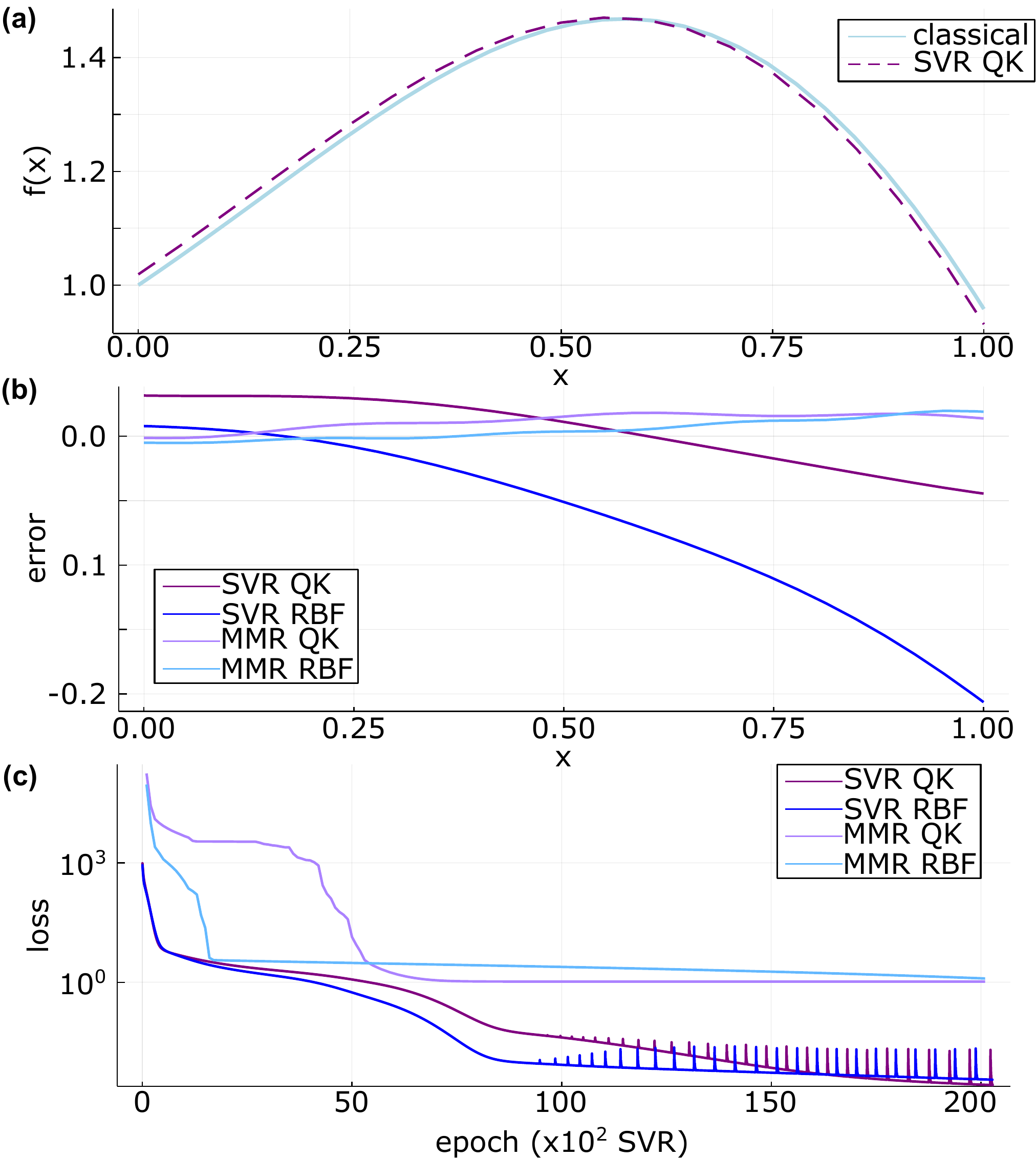}
\end{center}
\caption{MMR and SVR used to solve Duffing equation \eqref{eq:duffing} with $a =1, b=1, c=3, d=3, f(0) = 1$ and $f'(0) = 1$. (a) Solution via SVR method for two-layered quantum kernel with feature map of $R_x(qx/4)$ on each qubit with $q$ qubit index and $N=4$ (purple dashed line). Numerical solution from classical solver plotted with solid light blue curve. (b) Error between results and analytic solution plotted over $x$ as $f(x)-f_{\mathrm{num}}(x)$ (normalized by the range of values). Error plotted for result shown in (a) and SVR method with classical RBF kernel with $\sigma = 0.8$. Also plotted results from MMR method. The same form of kernels based on $N=8$ register, $\sigma = 0.8$. 
(c) Associated loss functions over epoch number for results shown in (a) and (b). SVR method uses 100 times more total epochs as compared to MMR.}
\label{fig:duffing}
\end{figure}

When implementing the MMR method we consider $\mathbf{x}$ and $\mathbf{y}$ of 13 points uniformly spaced over $[0,1]$. We use and compare the results from the classical RBF kernel ($\sigma = 0.2$ as before) and the quantum kernel with two layers of HEA circuits (depth equal to five) followed by feature map of $R_x (\phi(x))$ on each qubit with $\phi(x) = q x / 4$, where $q$ is the qubit index. As before, we consider eight qubits in the register. For the loss function MSE is used with a pinned boundary. This loss is optimized via Newtons method with 200 epochs. The error of the results is shown in Fig.~\ref{fig:duffing}(b) with associated loss functions in Fig.~\ref{fig:duffing}(c). As can be seen both kernel types are able to closely approximate the solution to the differential equation.

When implementing the SVR method the same $\mathbf{x}$ and simulation package are used. Different kernel functions are used however with less expressibility better suited to this method due to the differing function expression. Used are a classical kernel - the RBF kernel with $\sigma = 0.8$ and a quantum kernel - two layers of HEA depth five followed by feature map based on $R_x (\phi(x))$ on each qubit with $\phi(x) = q x / 4$, where $q$ is the qubit index and we consider four qubits in the register. $\gamma = 10^6$ is chosen. The resulting SVR system of equations to solve for a problem of the form $d^2f/dx^2 = g(x, f)$ with initial conditions $f(x_0) = f_0$ and $f'(x_0) = df_0$ is
\begin{align}
    \left[\begin{array}{c|c|c|c|c|c} \tilde{\Omega}_2^2 & \Omega_0^{2} & \mathbf{h}_0^2 & \mathbf{h}_0^2 & \mathbf{0} & \hat{0} \\
    \hline
    \Omega_2^{0} & \tilde{\Omega_0^0} & \mathbf{h}_0^0 & \mathbf{h}_1^0 & \mathbf{1} & -\hat{I} \\
    \hline
    {\mathbf{h}^T}_2^0 & {\mathbf{h}^T}_0^{0} & \tilde{h}_0^0 & \tilde{h}_1^0 & 1 & \mathbf{0}^T \\
    \hline
    {\mathbf{h}^T}_2^1 & {\mathbf{h}^T}_0^{1} & \tilde{h}_0^1 & \tilde{h}_1^1 & 0 & \mathbf{0}^T \\
    \hline
    \mathbf{0}^T & \mathbf{1}^T & 1 & 0 & 0 & \mathbf{0}^T \\
    \hline
    \hat{D} & \hat{I} & \mathbf{0} & \mathbf{0} & \mathbf{0} & \hat{0}
    \end{array}\right]
    \left[ \begin{array}{c} 
    \bm{\alpha} \\ \hline \bm{\eta} \\ \hline \beta_0 \\ \hline \beta_1 \\ \hline b \\ \hline \mathbf{y}
    \end{array} \right]= 
    \left[ \begin{array}{c} 
    \mathbf{\tilde{g}} \\ \hline \mathbf{0} \\ \hline f_0 \\ \hline df_0 \\ \hline 0 \\ \hline \hat{0}
    \end{array} \right],
\end{align}
where the notation is as follows:
\begin{align}
    [\Omega^m_n]_{i,j} &= \nabla^m_n \kappa(x_j, x_i), \\
    \tilde{\Omega}^m_n &= \Omega^m_n + \hat{I}/\gamma, \\
    [\mathbf{h}^m_n]_i &= \nabla^m_n \kappa(x_0, x_i), \\
    \tilde{h}^m_n &= \nabla^m_n \kappa(x_0, x_0) ,\\
    \hat{D} &= \mathrm{diag}\left( \left\{ \frac{\partial g}{\partial f}(x_i, y_i) \right \}_i \right),\\
    [\mathbf{\tilde{g}}]_i &= g(x_i, y_i) .
\end{align}

The resulting variational function is of the form $f(x) = \beta_0 \kappa(x_0, x) + \beta_1 \nabla_1^0 \kappa(x_0, x) + \sum_j \alpha_j \nabla_2^0 \kappa(x_j, x) + \sum_j \eta_j \kappa(x_j, x) + b$.
This system of nonlinear differential equations is trained using the ADAM optimizer with learning rate 0.003 and 20,000 epochs. The error of the results of this are shown in Fig.~\ref{fig:duffing}(b) with associated loss functions in Fig.~\ref{fig:duffing}(c). For this problem the quantum kernel is able to achieve a close solution, though not as close as MMR methods, the RBF kernel struggled however with maximum error magnitude over $0.2$.

\section{Discussion and Conclusion}

In this work, we proposed quantum protocols for solving differential equation with kernel methods. We represent potential solutions as quantum models that are based on weighted sums of kernel functions, corresponding to overlaps of quantum states. The adjustable weights are optimized such that for many problems the optimization is convex, leading to fast convergence to the potential solution. Specifically, we propose two approaches, being mixed model regression (MMR) and support vector regression (SVR), where optimization workflow is different. An important element of our approach is the automatic differentiation of quantum kernels with respect to encoded feature variables using quantum circuit differentiation. We applied both MMR and SVR for several toy problems. First, we presented regression for a quantum dataset, corresponding to nonequilibrium dynamics of quantum spin liquids. In this case, the use of quantum kernels may offer advantage, as native quantum operations are used. Second, we solve linear DEs, showing that nontrivial solutions can be routinely found with few epochs. Finally, applying our approaches to some nonlinear problems, the optimization becomes non-convex, thus requiring largely increased number of epochs. At the same time, we note that by kernelizing quantum models we modify the landscape of optimization. This raises the question of convergence difference between parameterized quantum circuits \cite{Larocca2021overparametrization,Larocca2021optimal} and kernel models. 

While this work presents a first step towards quantum kernel-based differential equation solving, many aspects are left unexplored. This includes the quantum feature map design, which should be chosen appropriately and could potentially be problem-motivated for each specific case. Finding conditions for which non-linear equations are guaranteed to result in a convex loss landscape is another open question.
Making use of a high-dimensional feature space without full tomography of the quantum wavefunction allows quantum kernel methods to potentially provide tangible advantage beyond classification.

Finally, let us discuss the comparison of quantum kernel-based approaches to solving DEs as compared to those based on differentiable quantum circuits \cite{kyriienko2021solving}, which in many ways reflect the difference between classical kernel methods and deep learning \cite{schuld2021quantum}. When considering the training stage, kernel methods require evaluating a Gram matrix of kernels with $O(|\mathbf{x}|^2)$ points, increasing measurement budget with the grid size $|\mathbf{x}|$ as compared to $O(|\mathbf{x}|)$ scaling of DQC evaluations for the loss function. At the same time, kernel methods do not rely on additional function evaluations, and optimization is straightforward for both linear and nonlinear problems. Deep learning instead needs to evaluate gradients at each iteration, leading to large overhead if the convergence is slow (for instance, when dealing with barren plateaus). This is a known trade-off for variational vs non-variational methods for ground state search \cite{Bespalova2021}. However, for trainable QML circuits it may be beneficial to do the iterative training, if the number of points in a dataset $|\mathbf{x}|$ is large. When considering model evaluation (reading out solution of DEs), kernel methods in principle require evaluating Gram matrix once again for a different grid, adding another $O(|\mathbf{x}|^2)$ computational steps. At the same time, deep learning and DQC only require evaluating the trained model at points of interest. Finding the optimum strategy between the two methods thus becomes crucially dependent on the problem and available quantum hardware. We expect that future studies will shed light on cases where one or another is preferred, both contributing to the emergent field of quantum DE solvers.

\textit{Ethics declaration.} A patent application for the method described in this manuscript has been submitted by Pasqal.


\begin{thebibliography}{81}
\expandafter\ifx\csname natexlab\endcsname\relax\def\natexlab#1{#1}\fi
\expandafter\ifx\csname bibnamefont\endcsname\relax
  \def\bibnamefont#1{#1}\fi
\expandafter\ifx\csname bibfnamefont\endcsname\relax
  \def\bibfnamefont#1{#1}\fi
\expandafter\ifx\csname citenamefont\endcsname\relax
  \def\citenamefont#1{#1}\fi
\expandafter\ifx\csname url\endcsname\relax
  \def\url#1{\texttt{#1}}\fi
\expandafter\ifx\csname urlprefix\endcsname\relax\def\urlprefix{URL }\fi
\providecommand{\bibinfo}[2]{#2}
\providecommand{\eprint}[2][]{\url{#2}}

\bibitem[{\citenamefont{Simmons}(2016)}]{simmons2016differential}
\bibinfo{author}{\bibfnamefont{G.~F.} \bibnamefont{Simmons}},
  \emph{\bibinfo{title}{Differential equations with applications and historical
  notes}} (\bibinfo{publisher}{CRC Press}, \bibinfo{year}{2016}).

\bibitem[{\citenamefont{Zachmanoglou and
  Thoe}(1986)}]{zachmanoglou1986introduction}
\bibinfo{author}{\bibfnamefont{E.~C.} \bibnamefont{Zachmanoglou}}
  \bibnamefont{and} \bibinfo{author}{\bibfnamefont{D.~W.} \bibnamefont{Thoe}},
  \emph{\bibinfo{title}{Introduction to partial differential equations with
  applications}} (\bibinfo{publisher}{Courier Corporation},
  \bibinfo{year}{1986}).

\bibitem[{\citenamefont{Smith et~al.}(1985)\citenamefont{Smith, Smith, and
  Smith}}]{smith1985numerical}
\bibinfo{author}{\bibfnamefont{G.~D.} \bibnamefont{Smith}},
  \bibinfo{author}{\bibfnamefont{G.~D.} \bibnamefont{Smith}}, \bibnamefont{and}
  \bibinfo{author}{\bibfnamefont{G.~D.~S.} \bibnamefont{Smith}},
  \emph{\bibinfo{title}{Numerical solution of partial differential equations:
  finite difference methods}} (\bibinfo{publisher}{Oxford university press},
  \bibinfo{year}{1985}).

\bibitem[{\citenamefont{Boyd}(2001)}]{boyd2001chebyshev}
\bibinfo{author}{\bibfnamefont{J.~P.} \bibnamefont{Boyd}},
  \emph{\bibinfo{title}{Chebyshev and Fourier spectral methods}}
  (\bibinfo{publisher}{Courier Corporation}, \bibinfo{year}{2001}).

\bibitem[{\citenamefont{Rackauckas and
  Nie}(2017)}]{rackauckas2017differentialequations}
\bibinfo{author}{\bibfnamefont{C.}~\bibnamefont{Rackauckas}} \bibnamefont{and}
  \bibinfo{author}{\bibfnamefont{Q.}~\bibnamefont{Nie}},
  \bibinfo{journal}{Journal of Open Research Software}
  \textbf{\bibinfo{volume}{5}}, \bibinfo{pages}{15} (\bibinfo{year}{2017}).

\bibitem[{\citenamefont{Rackauckas et~al.}(2019)\citenamefont{Rackauckas,
  Innes, Ma, Bettencourt, White, and Dixit}}]{rackauckas2019diffeqflux}
\bibinfo{author}{\bibfnamefont{C.}~\bibnamefont{Rackauckas}},
  \bibinfo{author}{\bibfnamefont{M.}~\bibnamefont{Innes}},
  \bibinfo{author}{\bibfnamefont{Y.}~\bibnamefont{Ma}},
  \bibinfo{author}{\bibfnamefont{J.}~\bibnamefont{Bettencourt}},
  \bibinfo{author}{\bibfnamefont{L.}~\bibnamefont{White}}, \bibnamefont{and}
  \bibinfo{author}{\bibfnamefont{V.}~\bibnamefont{Dixit}},
  \bibinfo{journal}{arXiv preprint arXiv:1902.02376}  (\bibinfo{year}{2019}).

\bibitem[{\citenamefont{Cai et~al.}(2022)\citenamefont{Cai, Mao, Wang, Yin, and
  Karniadakis}}]{Cai2022}
\bibinfo{author}{\bibfnamefont{S.}~\bibnamefont{Cai}},
  \bibinfo{author}{\bibfnamefont{Z.}~\bibnamefont{Mao}},
  \bibinfo{author}{\bibfnamefont{Z.}~\bibnamefont{Wang}},
  \bibinfo{author}{\bibfnamefont{M.}~\bibnamefont{Yin}}, \bibnamefont{and}
  \bibinfo{author}{\bibfnamefont{G.~E.} \bibnamefont{Karniadakis}},
  \bibinfo{journal}{Acta Mechanica Sinica}  (\bibinfo{year}{2022}), ISSN
  \bibinfo{issn}{1614-3116},
  \urlprefix\url{https://doi.org/10.1007/s10409-021-01148-1}.

\bibitem[{\citenamefont{Shor}(1994)}]{shor1994algorithms}
\bibinfo{author}{\bibfnamefont{P.~W.} \bibnamefont{Shor}}, in
  \emph{\bibinfo{booktitle}{Proceedings 35th annual symposium on foundations of
  computer science}} (\bibinfo{organization}{Ieee}, \bibinfo{year}{1994}), pp.
  \bibinfo{pages}{124--134}.

\bibitem[{\citenamefont{Harrow et~al.}(2009)\citenamefont{Harrow, Hassidim, and
  Lloyd}}]{harrow2009quantum}
\bibinfo{author}{\bibfnamefont{A.~W.} \bibnamefont{Harrow}},
  \bibinfo{author}{\bibfnamefont{A.}~\bibnamefont{Hassidim}}, \bibnamefont{and}
  \bibinfo{author}{\bibfnamefont{S.}~\bibnamefont{Lloyd}},
  \bibinfo{journal}{Physical review letters} \textbf{\bibinfo{volume}{103}},
  \bibinfo{pages}{150502} (\bibinfo{year}{2009}).

\bibitem[{\citenamefont{Arute et~al.}(2019)\citenamefont{Arute, Arya, Babbush,
  Bacon, Bardin, Barends, Biswas, Boixo, Brandao, Buell et~al.}}]{Arute2019}
\bibinfo{author}{\bibfnamefont{F.}~\bibnamefont{Arute}},
  \bibinfo{author}{\bibfnamefont{K.}~\bibnamefont{Arya}},
  \bibinfo{author}{\bibfnamefont{R.}~\bibnamefont{Babbush}},
  \bibinfo{author}{\bibfnamefont{D.}~\bibnamefont{Bacon}},
  \bibinfo{author}{\bibfnamefont{J.~C.} \bibnamefont{Bardin}},
  \bibinfo{author}{\bibfnamefont{R.}~\bibnamefont{Barends}},
  \bibinfo{author}{\bibfnamefont{R.}~\bibnamefont{Biswas}},
  \bibinfo{author}{\bibfnamefont{S.}~\bibnamefont{Boixo}},
  \bibinfo{author}{\bibfnamefont{F.~G. S.~L.} \bibnamefont{Brandao}},
  \bibinfo{author}{\bibfnamefont{D.~A.} \bibnamefont{Buell}},
  \bibnamefont{et~al.}, \bibinfo{journal}{Nature}
  \textbf{\bibinfo{volume}{574}}, \bibinfo{pages}{505} (\bibinfo{year}{2019}),
  ISSN \bibinfo{issn}{1476-4687},
  \urlprefix\url{https://doi.org/10.1038/s41586-019-1666-5}.

\bibitem[{\citenamefont{Nielsen and Chuang}(2000)}]{NielsenChuang}
\bibinfo{author}{\bibfnamefont{M.~A.} \bibnamefont{Nielsen}} \bibnamefont{and}
  \bibinfo{author}{\bibfnamefont{I.~L.} \bibnamefont{Chuang}},
  \emph{\bibinfo{title}{Quantum Computation and Quantum Information}}
  (\bibinfo{publisher}{Cambridge University Press}, \bibinfo{year}{2000}).

\bibitem[{\citenamefont{Biamonte et~al.}(2017)\citenamefont{Biamonte, Wittek,
  Pancotti, Rebentrost, Wiebe, and Lloyd}}]{Biamonte2017}
\bibinfo{author}{\bibfnamefont{J.}~\bibnamefont{Biamonte}},
  \bibinfo{author}{\bibfnamefont{P.}~\bibnamefont{Wittek}},
  \bibinfo{author}{\bibfnamefont{N.}~\bibnamefont{Pancotti}},
  \bibinfo{author}{\bibfnamefont{P.}~\bibnamefont{Rebentrost}},
  \bibinfo{author}{\bibfnamefont{N.}~\bibnamefont{Wiebe}}, \bibnamefont{and}
  \bibinfo{author}{\bibfnamefont{S.}~\bibnamefont{Lloyd}},
  \bibinfo{journal}{Nature} \textbf{\bibinfo{volume}{549}},
  \bibinfo{pages}{195} (\bibinfo{year}{2017}), ISSN \bibinfo{issn}{1476-4687},
  \urlprefix\url{https://doi.org/10.1038/nature23474}.

\bibitem[{\citenamefont{Leyton and Osborne}(2008)}]{leyton2008quantum}
\bibinfo{author}{\bibfnamefont{S.~K.} \bibnamefont{Leyton}} \bibnamefont{and}
  \bibinfo{author}{\bibfnamefont{T.~J.} \bibnamefont{Osborne}},
  \bibinfo{journal}{arXiv preprint arXiv:0812.4423}  (\bibinfo{year}{2008}).

\bibitem[{\citenamefont{Berry et~al.}(2017)\citenamefont{Berry, Childs,
  Ostrander, and Wang}}]{Berry2017}
\bibinfo{author}{\bibfnamefont{D.~W.} \bibnamefont{Berry}},
  \bibinfo{author}{\bibfnamefont{A.~M.} \bibnamefont{Childs}},
  \bibinfo{author}{\bibfnamefont{A.}~\bibnamefont{Ostrander}},
  \bibnamefont{and} \bibinfo{author}{\bibfnamefont{G.}~\bibnamefont{Wang}},
  \bibinfo{journal}{Communications in Mathematical Physics}
  \textbf{\bibinfo{volume}{356}}, \bibinfo{pages}{1057} (\bibinfo{year}{2017}),
  ISSN \bibinfo{issn}{1432-0916},
  \urlprefix\url{https://doi.org/10.1007/s00220-017-3002-y}.

\bibitem[{\citenamefont{Lloyd et~al.}(2020)\citenamefont{Lloyd, Palma, Gokler,
  Kiani, Liu, Marvian, Tennie, and Palmer}}]{lloyd2020quantum}
\bibinfo{author}{\bibfnamefont{S.}~\bibnamefont{Lloyd}},
  \bibinfo{author}{\bibfnamefont{G.~D.} \bibnamefont{Palma}},
  \bibinfo{author}{\bibfnamefont{C.}~\bibnamefont{Gokler}},
  \bibinfo{author}{\bibfnamefont{B.}~\bibnamefont{Kiani}},
  \bibinfo{author}{\bibfnamefont{Z.-W.} \bibnamefont{Liu}},
  \bibinfo{author}{\bibfnamefont{M.}~\bibnamefont{Marvian}},
  \bibinfo{author}{\bibfnamefont{F.}~\bibnamefont{Tennie}}, \bibnamefont{and}
  \bibinfo{author}{\bibfnamefont{T.}~\bibnamefont{Palmer}},
  \emph{\bibinfo{title}{Quantum algorithm for nonlinear differential
  equations}} (\bibinfo{year}{2020}), \eprint{2011.06571}.

\bibitem[{\citenamefont{Liu et~al.}(2021)\citenamefont{Liu, Øie Kolden, Krovi,
  Loureiro, Trivisa, and Childs}}]{LiuChilds2021}
\bibinfo{author}{\bibfnamefont{J.-P.} \bibnamefont{Liu}},
  \bibinfo{author}{\bibfnamefont{H.}~\bibnamefont{Øie Kolden}},
  \bibinfo{author}{\bibfnamefont{H.~K.} \bibnamefont{Krovi}},
  \bibinfo{author}{\bibfnamefont{N.~F.} \bibnamefont{Loureiro}},
  \bibinfo{author}{\bibfnamefont{K.}~\bibnamefont{Trivisa}}, \bibnamefont{and}
  \bibinfo{author}{\bibfnamefont{A.~M.} \bibnamefont{Childs}},
  \bibinfo{journal}{Proceedings of the National Academy of Sciences}
  \textbf{\bibinfo{volume}{118}}, \bibinfo{pages}{e2026805118}
  (\bibinfo{year}{2021}),
  \eprint{https://www.pnas.org/doi/pdf/10.1073/pnas.2026805118},
  \urlprefix\url{https://www.pnas.org/doi/abs/10.1073/pnas.2026805118}.

\bibitem[{\citenamefont{Jin and Liu}(2022)}]{Jin2022}
\bibinfo{author}{\bibfnamefont{S.}~\bibnamefont{Jin}} \bibnamefont{and}
  \bibinfo{author}{\bibfnamefont{N.}~\bibnamefont{Liu}} (\bibinfo{year}{2022}),
  \eprint{2202.07834}, \urlprefix\url{https://arxiv.org/abs/2202.07834}.

\bibitem[{\citenamefont{Linden et~al.}(2020)\citenamefont{Linden, Montanaro,
  and Shao}}]{Linden2020}
\bibinfo{author}{\bibfnamefont{N.}~\bibnamefont{Linden}},
  \bibinfo{author}{\bibfnamefont{A.}~\bibnamefont{Montanaro}},
  \bibnamefont{and} \bibinfo{author}{\bibfnamefont{C.}~\bibnamefont{Shao}}
  (\bibinfo{year}{2020}), \eprint{2004.06516},
  \urlprefix\url{http://arxiv.org/abs/2004.06516}.

\bibitem[{\citenamefont{Scherer et~al.}(2017)\citenamefont{Scherer, Valiron,
  Mau, Alexander, van den Berg, and Chapuran}}]{Scherer2017}
\bibinfo{author}{\bibfnamefont{A.}~\bibnamefont{Scherer}},
  \bibinfo{author}{\bibfnamefont{B.}~\bibnamefont{Valiron}},
  \bibinfo{author}{\bibfnamefont{S.-C.} \bibnamefont{Mau}},
  \bibinfo{author}{\bibfnamefont{S.}~\bibnamefont{Alexander}},
  \bibinfo{author}{\bibfnamefont{E.}~\bibnamefont{van den Berg}},
  \bibnamefont{and} \bibinfo{author}{\bibfnamefont{T.~E.}
  \bibnamefont{Chapuran}}, \bibinfo{journal}{Quantum Information Processing}
  \textbf{\bibinfo{volume}{16}}, \bibinfo{pages}{60} (\bibinfo{year}{2017}),
  ISSN \bibinfo{issn}{1573-1332},
  \urlprefix\url{https://doi.org/10.1007/s11128-016-1495-5}.

\bibitem[{\citenamefont{Bharti et~al.}(2022)\citenamefont{Bharti,
  Cervera-Lierta, Kyaw, Haug, Alperin-Lea, Anand, Degroote, Heimonen, Kottmann,
  Menke et~al.}}]{Bharti2022}
\bibinfo{author}{\bibfnamefont{K.}~\bibnamefont{Bharti}},
  \bibinfo{author}{\bibfnamefont{A.}~\bibnamefont{Cervera-Lierta}},
  \bibinfo{author}{\bibfnamefont{T.~H.} \bibnamefont{Kyaw}},
  \bibinfo{author}{\bibfnamefont{T.}~\bibnamefont{Haug}},
  \bibinfo{author}{\bibfnamefont{S.}~\bibnamefont{Alperin-Lea}},
  \bibinfo{author}{\bibfnamefont{A.}~\bibnamefont{Anand}},
  \bibinfo{author}{\bibfnamefont{M.}~\bibnamefont{Degroote}},
  \bibinfo{author}{\bibfnamefont{H.}~\bibnamefont{Heimonen}},
  \bibinfo{author}{\bibfnamefont{J.~S.} \bibnamefont{Kottmann}},
  \bibinfo{author}{\bibfnamefont{T.}~\bibnamefont{Menke}},
  \bibnamefont{et~al.}, \bibinfo{journal}{Rev. Mod. Phys.}
  \textbf{\bibinfo{volume}{94}}, \bibinfo{pages}{015004}
  (\bibinfo{year}{2022}),
  \urlprefix\url{https://link.aps.org/doi/10.1103/RevModPhys.94.015004}.

\bibitem[{\citenamefont{Cerezo et~al.}(2021)\citenamefont{Cerezo, Arrasmith,
  Babbush, Benjamin, Endo, Fujii, McClean, Mitarai, Yuan, Cincio
  et~al.}}]{cerezo2021variational}
\bibinfo{author}{\bibfnamefont{M.}~\bibnamefont{Cerezo}},
  \bibinfo{author}{\bibfnamefont{A.}~\bibnamefont{Arrasmith}},
  \bibinfo{author}{\bibfnamefont{R.}~\bibnamefont{Babbush}},
  \bibinfo{author}{\bibfnamefont{S.~C.} \bibnamefont{Benjamin}},
  \bibinfo{author}{\bibfnamefont{S.}~\bibnamefont{Endo}},
  \bibinfo{author}{\bibfnamefont{K.}~\bibnamefont{Fujii}},
  \bibinfo{author}{\bibfnamefont{J.~R.} \bibnamefont{McClean}},
  \bibinfo{author}{\bibfnamefont{K.}~\bibnamefont{Mitarai}},
  \bibinfo{author}{\bibfnamefont{X.}~\bibnamefont{Yuan}},
  \bibinfo{author}{\bibfnamefont{L.}~\bibnamefont{Cincio}},
  \bibnamefont{et~al.}, \bibinfo{journal}{Nature Reviews Physics}
  \textbf{\bibinfo{volume}{3}}, \bibinfo{pages}{625} (\bibinfo{year}{2021}).

\bibitem[{\citenamefont{Benedetti et~al.}(2019)\citenamefont{Benedetti, Lloyd,
  Sack, and Fiorentini}}]{Benedetti2019}
\bibinfo{author}{\bibfnamefont{M.}~\bibnamefont{Benedetti}},
  \bibinfo{author}{\bibfnamefont{E.}~\bibnamefont{Lloyd}},
  \bibinfo{author}{\bibfnamefont{S.}~\bibnamefont{Sack}}, \bibnamefont{and}
  \bibinfo{author}{\bibfnamefont{M.}~\bibnamefont{Fiorentini}},
  \bibinfo{journal}{Quantum Science and Technology}
  \textbf{\bibinfo{volume}{4}}, \bibinfo{pages}{043001} (\bibinfo{year}{2019}),
  \urlprefix\url{https://doi.org/10.1088/2058-9565/ab4eb5}.

\bibitem[{\citenamefont{Perdomo-Ortiz et~al.}(2018)\citenamefont{Perdomo-Ortiz,
  Benedetti, Realpe-G{\'{o}}mez, and Biswas}}]{PerdomoOrtiz2018}
\bibinfo{author}{\bibfnamefont{A.}~\bibnamefont{Perdomo-Ortiz}},
  \bibinfo{author}{\bibfnamefont{M.}~\bibnamefont{Benedetti}},
  \bibinfo{author}{\bibfnamefont{J.}~\bibnamefont{Realpe-G{\'{o}}mez}},
  \bibnamefont{and} \bibinfo{author}{\bibfnamefont{R.}~\bibnamefont{Biswas}},
  \bibinfo{journal}{Quantum Science and Technology}
  \textbf{\bibinfo{volume}{3}}, \bibinfo{pages}{030502} (\bibinfo{year}{2018}),
  \urlprefix\url{https://doi.org/10.1088/2058-9565/aab859}.

\bibitem[{\citenamefont{Schuld and Killoran}(2019)}]{Schuld2019QML}
\bibinfo{author}{\bibfnamefont{M.}~\bibnamefont{Schuld}} \bibnamefont{and}
  \bibinfo{author}{\bibfnamefont{N.}~\bibnamefont{Killoran}},
  \bibinfo{journal}{Phys. Rev. Lett.} \textbf{\bibinfo{volume}{122}},
  \bibinfo{pages}{040504} (\bibinfo{year}{2019}),
  \urlprefix\url{https://link.aps.org/doi/10.1103/PhysRevLett.122.040504}.

\bibitem[{\citenamefont{Mitarai et~al.}(2018)\citenamefont{Mitarai, Negoro,
  Kitagawa, and Fujii}}]{mitarai2018quantum}
\bibinfo{author}{\bibfnamefont{K.}~\bibnamefont{Mitarai}},
  \bibinfo{author}{\bibfnamefont{M.}~\bibnamefont{Negoro}},
  \bibinfo{author}{\bibfnamefont{M.}~\bibnamefont{Kitagawa}}, \bibnamefont{and}
  \bibinfo{author}{\bibfnamefont{K.}~\bibnamefont{Fujii}},
  \bibinfo{journal}{Physical Review A} \textbf{\bibinfo{volume}{98}},
  \bibinfo{pages}{032309} (\bibinfo{year}{2018}).

\bibitem[{\citenamefont{Liu and Wang}(2018)}]{Liu2018}
\bibinfo{author}{\bibfnamefont{J.-G.} \bibnamefont{Liu}} \bibnamefont{and}
  \bibinfo{author}{\bibfnamefont{L.}~\bibnamefont{Wang}},
  \bibinfo{journal}{Phys. Rev. A} \textbf{\bibinfo{volume}{98}},
  \bibinfo{pages}{062324} (\bibinfo{year}{2018}),
  \urlprefix\url{https://link.aps.org/doi/10.1103/PhysRevA.98.062324}.

\bibitem[{\citenamefont{Zoufal et~al.}(2019)\citenamefont{Zoufal, Lucchi, and
  Woerner}}]{Zoufal2019}
\bibinfo{author}{\bibfnamefont{C.}~\bibnamefont{Zoufal}},
  \bibinfo{author}{\bibfnamefont{A.}~\bibnamefont{Lucchi}}, \bibnamefont{and}
  \bibinfo{author}{\bibfnamefont{S.}~\bibnamefont{Woerner}},
  \bibinfo{journal}{npj Quantum Information} \textbf{\bibinfo{volume}{5}},
  \bibinfo{pages}{103} (\bibinfo{year}{2019}), ISSN \bibinfo{issn}{2056-6387},
  \urlprefix\url{https://doi.org/10.1038/s41534-019-0223-2}.

\bibitem[{\citenamefont{Coyle et~al.}(2020)\citenamefont{Coyle, Mills, Danos,
  and Kashefi}}]{Coyle2020}
\bibinfo{author}{\bibfnamefont{B.}~\bibnamefont{Coyle}},
  \bibinfo{author}{\bibfnamefont{D.}~\bibnamefont{Mills}},
  \bibinfo{author}{\bibfnamefont{V.}~\bibnamefont{Danos}}, \bibnamefont{and}
  \bibinfo{author}{\bibfnamefont{E.}~\bibnamefont{Kashefi}},
  \bibinfo{journal}{npj Quantum Information} \textbf{\bibinfo{volume}{6}},
  \bibinfo{pages}{60} (\bibinfo{year}{2020}), ISSN \bibinfo{issn}{2056-6387},
  \urlprefix\url{https://doi.org/10.1038/s41534-020-00288-9}.

\bibitem[{\citenamefont{Abbas et~al.}(2021)\citenamefont{Abbas, Sutter, Zoufal,
  Lucchi, Figalli, and Woerner}}]{Abbas2021}
\bibinfo{author}{\bibfnamefont{A.}~\bibnamefont{Abbas}},
  \bibinfo{author}{\bibfnamefont{D.}~\bibnamefont{Sutter}},
  \bibinfo{author}{\bibfnamefont{C.}~\bibnamefont{Zoufal}},
  \bibinfo{author}{\bibfnamefont{A.}~\bibnamefont{Lucchi}},
  \bibinfo{author}{\bibfnamefont{A.}~\bibnamefont{Figalli}}, \bibnamefont{and}
  \bibinfo{author}{\bibfnamefont{S.}~\bibnamefont{Woerner}},
  \bibinfo{journal}{Nature Computational Science} \textbf{\bibinfo{volume}{1}},
  \bibinfo{pages}{403} (\bibinfo{year}{2021}), ISSN \bibinfo{issn}{2662-8457},
  \urlprefix\url{https://doi.org/10.1038/s43588-021-00084-1}.

\bibitem[{\citenamefont{Du et~al.}(2021)\citenamefont{Du, Hsieh, Liu, You, and
  Tao}}]{Du2021}
\bibinfo{author}{\bibfnamefont{Y.}~\bibnamefont{Du}},
  \bibinfo{author}{\bibfnamefont{M.-H.} \bibnamefont{Hsieh}},
  \bibinfo{author}{\bibfnamefont{T.}~\bibnamefont{Liu}},
  \bibinfo{author}{\bibfnamefont{S.}~\bibnamefont{You}}, \bibnamefont{and}
  \bibinfo{author}{\bibfnamefont{D.}~\bibnamefont{Tao}}, \bibinfo{journal}{PRX
  Quantum} \textbf{\bibinfo{volume}{2}}, \bibinfo{pages}{040337}
  (\bibinfo{year}{2021}),
  \urlprefix\url{https://link.aps.org/doi/10.1103/PRXQuantum.2.040337}.

\bibitem[{\citenamefont{Huang et~al.}(2021{\natexlab{a}})\citenamefont{Huang,
  Du, Gong, Zhao, Wu, Wang, Li, Liang, Lin, Xu et~al.}}]{Huang2021PRAppl}
\bibinfo{author}{\bibfnamefont{H.-L.} \bibnamefont{Huang}},
  \bibinfo{author}{\bibfnamefont{Y.}~\bibnamefont{Du}},
  \bibinfo{author}{\bibfnamefont{M.}~\bibnamefont{Gong}},
  \bibinfo{author}{\bibfnamefont{Y.}~\bibnamefont{Zhao}},
  \bibinfo{author}{\bibfnamefont{Y.}~\bibnamefont{Wu}},
  \bibinfo{author}{\bibfnamefont{C.}~\bibnamefont{Wang}},
  \bibinfo{author}{\bibfnamefont{S.}~\bibnamefont{Li}},
  \bibinfo{author}{\bibfnamefont{F.}~\bibnamefont{Liang}},
  \bibinfo{author}{\bibfnamefont{J.}~\bibnamefont{Lin}},
  \bibinfo{author}{\bibfnamefont{Y.}~\bibnamefont{Xu}}, \bibnamefont{et~al.},
  \bibinfo{journal}{Phys. Rev. Applied} \textbf{\bibinfo{volume}{16}},
  \bibinfo{pages}{024051} (\bibinfo{year}{2021}{\natexlab{a}}),
  \urlprefix\url{https://link.aps.org/doi/10.1103/PhysRevApplied.16.024051}.

\bibitem[{\citenamefont{Chen et~al.}(2020)\citenamefont{Chen, Yang, Qi, Chen,
  Ma, and Goan}}]{SChen2020a}
\bibinfo{author}{\bibfnamefont{S.~Y.-C.} \bibnamefont{Chen}},
  \bibinfo{author}{\bibfnamefont{C.-H.~H.} \bibnamefont{Yang}},
  \bibinfo{author}{\bibfnamefont{J.}~\bibnamefont{Qi}},
  \bibinfo{author}{\bibfnamefont{P.-Y.} \bibnamefont{Chen}},
  \bibinfo{author}{\bibfnamefont{X.}~\bibnamefont{Ma}}, \bibnamefont{and}
  \bibinfo{author}{\bibfnamefont{H.-S.} \bibnamefont{Goan}},
  \bibinfo{journal}{IEEE Access} \textbf{\bibinfo{volume}{8}},
  \bibinfo{pages}{141007} (\bibinfo{year}{2020}).

\bibitem[{\citenamefont{Wu et~al.}(2021)\citenamefont{Wu, Chan, Guan, Sun,
  Wang, Zhou, Livny, Carminati, Meglio, Li et~al.}}]{Wu2021}
\bibinfo{author}{\bibfnamefont{S.~L.} \bibnamefont{Wu}},
  \bibinfo{author}{\bibfnamefont{J.}~\bibnamefont{Chan}},
  \bibinfo{author}{\bibfnamefont{W.}~\bibnamefont{Guan}},
  \bibinfo{author}{\bibfnamefont{S.}~\bibnamefont{Sun}},
  \bibinfo{author}{\bibfnamefont{A.}~\bibnamefont{Wang}},
  \bibinfo{author}{\bibfnamefont{C.}~\bibnamefont{Zhou}},
  \bibinfo{author}{\bibfnamefont{M.}~\bibnamefont{Livny}},
  \bibinfo{author}{\bibfnamefont{F.}~\bibnamefont{Carminati}},
  \bibinfo{author}{\bibfnamefont{A.~D.} \bibnamefont{Meglio}},
  \bibinfo{author}{\bibfnamefont{A.~C.~Y.} \bibnamefont{Li}},
  \bibnamefont{et~al.}, \bibinfo{journal}{Journal of Physics G: Nuclear and
  Particle Physics} \textbf{\bibinfo{volume}{48}}, \bibinfo{pages}{125003}
  (\bibinfo{year}{2021}),
  \urlprefix\url{https://doi.org/10.1088/1361-6471/ac1391}.

\bibitem[{\citenamefont{Chen and Yoo}(2021)}]{SChen2021}
\bibinfo{author}{\bibfnamefont{S.~Y.-C.} \bibnamefont{Chen}} \bibnamefont{and}
  \bibinfo{author}{\bibfnamefont{S.}~\bibnamefont{Yoo}},
  \bibinfo{journal}{Entropy} \textbf{\bibinfo{volume}{23}}
  (\bibinfo{year}{2021}), ISSN \bibinfo{issn}{1099-4300},
  \urlprefix\url{https://www.mdpi.com/1099-4300/23/4/460}.

\bibitem[{\citenamefont{Endo et~al.}(2020)\citenamefont{Endo, Sun, Li,
  Benjamin, and Yuan}}]{Endo2020}
\bibinfo{author}{\bibfnamefont{S.}~\bibnamefont{Endo}},
  \bibinfo{author}{\bibfnamefont{J.}~\bibnamefont{Sun}},
  \bibinfo{author}{\bibfnamefont{Y.}~\bibnamefont{Li}},
  \bibinfo{author}{\bibfnamefont{S.~C.} \bibnamefont{Benjamin}},
  \bibnamefont{and} \bibinfo{author}{\bibfnamefont{X.}~\bibnamefont{Yuan}},
  \bibinfo{journal}{Phys. Rev. Lett.} \textbf{\bibinfo{volume}{125}},
  \bibinfo{pages}{010501} (\bibinfo{year}{2020}),
  \urlprefix\url{https://link.aps.org/doi/10.1103/PhysRevLett.125.010501}.

\bibitem[{\citenamefont{C{\^i}rstoiu et~al.}(2020)\citenamefont{C{\^i}rstoiu,
  Holmes, Iosue, Cincio, Coles, and Sornborger}}]{Cirstoiu2020}
\bibinfo{author}{\bibfnamefont{C.}~\bibnamefont{C{\^i}rstoiu}},
  \bibinfo{author}{\bibfnamefont{Z.}~\bibnamefont{Holmes}},
  \bibinfo{author}{\bibfnamefont{J.}~\bibnamefont{Iosue}},
  \bibinfo{author}{\bibfnamefont{L.}~\bibnamefont{Cincio}},
  \bibinfo{author}{\bibfnamefont{P.~J.} \bibnamefont{Coles}}, \bibnamefont{and}
  \bibinfo{author}{\bibfnamefont{A.}~\bibnamefont{Sornborger}},
  \bibinfo{journal}{npj Quantum Information} \textbf{\bibinfo{volume}{6}},
  \bibinfo{pages}{82} (\bibinfo{year}{2020}), ISSN \bibinfo{issn}{2056-6387},
  \urlprefix\url{https://doi.org/10.1038/s41534-020-00302-0}.

\bibitem[{\citenamefont{Xu et~al.}(2021)\citenamefont{Xu, Sun, Endo, Li,
  Benjamin, and Yuan}}]{XU20212181}
\bibinfo{author}{\bibfnamefont{X.}~\bibnamefont{Xu}},
  \bibinfo{author}{\bibfnamefont{J.}~\bibnamefont{Sun}},
  \bibinfo{author}{\bibfnamefont{S.}~\bibnamefont{Endo}},
  \bibinfo{author}{\bibfnamefont{Y.}~\bibnamefont{Li}},
  \bibinfo{author}{\bibfnamefont{S.~C.} \bibnamefont{Benjamin}},
  \bibnamefont{and} \bibinfo{author}{\bibfnamefont{X.}~\bibnamefont{Yuan}},
  \bibinfo{journal}{Science Bulletin} \textbf{\bibinfo{volume}{66}},
  \bibinfo{pages}{2181} (\bibinfo{year}{2021}), ISSN \bibinfo{issn}{2095-9273},
  \urlprefix\url{https://www.sciencedirect.com/science/article/pii/S2095927321004631}.

\bibitem[{\citenamefont{Bravo-Prieto et~al.}(2019)\citenamefont{Bravo-Prieto,
  LaRose, Cerezo, Subasi, Cincio, and Coles}}]{Bravo-Prieto2019}
\bibinfo{author}{\bibfnamefont{C.}~\bibnamefont{Bravo-Prieto}},
  \bibinfo{author}{\bibfnamefont{R.}~\bibnamefont{LaRose}},
  \bibinfo{author}{\bibfnamefont{M.}~\bibnamefont{Cerezo}},
  \bibinfo{author}{\bibfnamefont{Y.}~\bibnamefont{Subasi}},
  \bibinfo{author}{\bibfnamefont{L.}~\bibnamefont{Cincio}}, \bibnamefont{and}
  \bibinfo{author}{\bibfnamefont{P.~J.} \bibnamefont{Coles}}
  (\bibinfo{year}{2019}), \eprint{1909.05820},
  \urlprefix\url{http://arxiv.org/abs/1909.05820}.

\bibitem[{\citenamefont{Chen et~al.}(2019)\citenamefont{Chen, Shiau, Wu, and
  Wu}}]{Chen2019}
\bibinfo{author}{\bibfnamefont{C.-C.} \bibnamefont{Chen}},
  \bibinfo{author}{\bibfnamefont{S.-Y.} \bibnamefont{Shiau}},
  \bibinfo{author}{\bibfnamefont{M.-F.} \bibnamefont{Wu}}, \bibnamefont{and}
  \bibinfo{author}{\bibfnamefont{Y.-R.} \bibnamefont{Wu}},
  \bibinfo{journal}{Scientific Reports} \textbf{\bibinfo{volume}{9}},
  \bibinfo{pages}{16251} (\bibinfo{year}{2019}), ISSN
  \bibinfo{issn}{2045-2322},
  \urlprefix\url{https://doi.org/10.1038/s41598-019-52275-6}.

\bibitem[{\citenamefont{Lubasch et~al.}(2020)\citenamefont{Lubasch, Joo,
  Moinier, Kiffner, and Jaksch}}]{lubasch2020variational}
\bibinfo{author}{\bibfnamefont{M.}~\bibnamefont{Lubasch}},
  \bibinfo{author}{\bibfnamefont{J.}~\bibnamefont{Joo}},
  \bibinfo{author}{\bibfnamefont{P.}~\bibnamefont{Moinier}},
  \bibinfo{author}{\bibfnamefont{M.}~\bibnamefont{Kiffner}}, \bibnamefont{and}
  \bibinfo{author}{\bibfnamefont{D.}~\bibnamefont{Jaksch}},
  \bibinfo{journal}{Physical Review A} \textbf{\bibinfo{volume}{101}},
  \bibinfo{pages}{010301} (\bibinfo{year}{2020}).

\bibitem[{\citenamefont{Kyriienko et~al.}(2021)\citenamefont{Kyriienko, Paine,
  and Elfving}}]{kyriienko2021solving}
\bibinfo{author}{\bibfnamefont{O.}~\bibnamefont{Kyriienko}},
  \bibinfo{author}{\bibfnamefont{A.~E.} \bibnamefont{Paine}}, \bibnamefont{and}
  \bibinfo{author}{\bibfnamefont{V.~E.} \bibnamefont{Elfving}},
  \bibinfo{journal}{Physical Review A} \textbf{\bibinfo{volume}{103}},
  \bibinfo{pages}{052416} (\bibinfo{year}{2021}).

\bibitem[{\citenamefont{Schuld et~al.}(2021)\citenamefont{Schuld, Sweke, and
  Meyer}}]{SchuldSweke2021}
\bibinfo{author}{\bibfnamefont{M.}~\bibnamefont{Schuld}},
  \bibinfo{author}{\bibfnamefont{R.}~\bibnamefont{Sweke}}, \bibnamefont{and}
  \bibinfo{author}{\bibfnamefont{J.~J.} \bibnamefont{Meyer}},
  \bibinfo{journal}{Phys. Rev. A} \textbf{\bibinfo{volume}{103}},
  \bibinfo{pages}{032430} (\bibinfo{year}{2021}),
  \urlprefix\url{https://link.aps.org/doi/10.1103/PhysRevA.103.032430}.

\bibitem[{\citenamefont{Schuld et~al.}(2019)\citenamefont{Schuld, Bergholm,
  Gogolin, Izaac, and Killoran}}]{schuld2019evaluating}
\bibinfo{author}{\bibfnamefont{M.}~\bibnamefont{Schuld}},
  \bibinfo{author}{\bibfnamefont{V.}~\bibnamefont{Bergholm}},
  \bibinfo{author}{\bibfnamefont{C.}~\bibnamefont{Gogolin}},
  \bibinfo{author}{\bibfnamefont{J.}~\bibnamefont{Izaac}}, \bibnamefont{and}
  \bibinfo{author}{\bibfnamefont{N.}~\bibnamefont{Killoran}},
  \bibinfo{journal}{Physical Review A} \textbf{\bibinfo{volume}{99}},
  \bibinfo{pages}{032331} (\bibinfo{year}{2019}).

\bibitem[{\citenamefont{Knudsen and Mendl}(2020)}]{Knudsen2020}
\bibinfo{author}{\bibfnamefont{M.}~\bibnamefont{Knudsen}} \bibnamefont{and}
  \bibinfo{author}{\bibfnamefont{C.~B.} \bibnamefont{Mendl}}
  (\bibinfo{year}{2020}), \eprint{2012.12220},
  \urlprefix\url{https://arxiv.org/abs/2012.12220}.

\bibitem[{\citenamefont{Paine et~al.}(2021)\citenamefont{Paine, Elfving, and
  Kyriienko}}]{Paine2021}
\bibinfo{author}{\bibfnamefont{A.~E.} \bibnamefont{Paine}},
  \bibinfo{author}{\bibfnamefont{V.~E.} \bibnamefont{Elfving}},
  \bibnamefont{and} \bibinfo{author}{\bibfnamefont{O.}~\bibnamefont{Kyriienko}}
  (\bibinfo{year}{2021}), \eprint{2108.03190},
  \urlprefix\url{https://arxiv.org/abs/2108.03190}.

\bibitem[{\citenamefont{Romero and Aspuru-Guzik}(2021)}]{Romero2021}
\bibinfo{author}{\bibfnamefont{J.}~\bibnamefont{Romero}} \bibnamefont{and}
  \bibinfo{author}{\bibfnamefont{A.}~\bibnamefont{Aspuru-Guzik}},
  \bibinfo{journal}{Advanced Quantum Technologies}
  \textbf{\bibinfo{volume}{4}}, \bibinfo{pages}{2000003}
  (\bibinfo{year}{2021}),
  \eprint{https://onlinelibrary.wiley.com/doi/pdf/10.1002/qute.202000003},
  \urlprefix\url{https://onlinelibrary.wiley.com/doi/abs/10.1002/qute.202000003}.

\bibitem[{\citenamefont{Kyriienko et~al.}(2022)\citenamefont{Kyriienko, Paine,
  and Elfving}}]{Kyriienko2022}
\bibinfo{author}{\bibfnamefont{O.}~\bibnamefont{Kyriienko}},
  \bibinfo{author}{\bibfnamefont{A.~E.} \bibnamefont{Paine}}, \bibnamefont{and}
  \bibinfo{author}{\bibfnamefont{V.~E.} \bibnamefont{Elfving}}
  (\bibinfo{year}{2022}), \eprint{2202.08253},
  \urlprefix\url{https://arxiv.org/abs/2202.08253}.

\bibitem[{\citenamefont{Havl{\'i}{\v{c}}ek
  et~al.}(2019)\citenamefont{Havl{\'i}{\v{c}}ek, C{\'o}rcoles, Temme, Harrow,
  Kandala, Chow, and Gambetta}}]{Havlicek2019}
\bibinfo{author}{\bibfnamefont{V.}~\bibnamefont{Havl{\'i}{\v{c}}ek}},
  \bibinfo{author}{\bibfnamefont{A.~D.} \bibnamefont{C{\'o}rcoles}},
  \bibinfo{author}{\bibfnamefont{K.}~\bibnamefont{Temme}},
  \bibinfo{author}{\bibfnamefont{A.~W.} \bibnamefont{Harrow}},
  \bibinfo{author}{\bibfnamefont{A.}~\bibnamefont{Kandala}},
  \bibinfo{author}{\bibfnamefont{J.~M.} \bibnamefont{Chow}}, \bibnamefont{and}
  \bibinfo{author}{\bibfnamefont{J.~M.} \bibnamefont{Gambetta}},
  \bibinfo{journal}{Nature} \textbf{\bibinfo{volume}{567}},
  \bibinfo{pages}{209} (\bibinfo{year}{2019}), ISSN \bibinfo{issn}{1476-4687},
  \urlprefix\url{https://doi.org/10.1038/s41586-019-0980-2}.

\bibitem[{\citenamefont{Ardeshir et~al.}(2021)\citenamefont{Ardeshir, Sanford,
  and Hsu}}]{ardeshir2021support}
\bibinfo{author}{\bibfnamefont{N.}~\bibnamefont{Ardeshir}},
  \bibinfo{author}{\bibfnamefont{C.}~\bibnamefont{Sanford}}, \bibnamefont{and}
  \bibinfo{author}{\bibfnamefont{D.~J.} \bibnamefont{Hsu}},
  \bibinfo{journal}{Advances in Neural Information Processing Systems}
  \textbf{\bibinfo{volume}{34}} (\bibinfo{year}{2021}).

\bibitem[{\citenamefont{Kung}(2014)}]{kung2014kernel}
\bibinfo{author}{\bibfnamefont{S.~Y.} \bibnamefont{Kung}},
  \emph{\bibinfo{title}{Kernel methods and machine learning}}
  (\bibinfo{publisher}{Cambridge University Press}, \bibinfo{year}{2014}).

\bibitem[{\citenamefont{Huang et~al.}(2021{\natexlab{b}})\citenamefont{Huang,
  Broughton, Mohseni, Babbush, Boixo, Neven, and McClean}}]{Huang2021}
\bibinfo{author}{\bibfnamefont{H.-Y.} \bibnamefont{Huang}},
  \bibinfo{author}{\bibfnamefont{M.}~\bibnamefont{Broughton}},
  \bibinfo{author}{\bibfnamefont{M.}~\bibnamefont{Mohseni}},
  \bibinfo{author}{\bibfnamefont{R.}~\bibnamefont{Babbush}},
  \bibinfo{author}{\bibfnamefont{S.}~\bibnamefont{Boixo}},
  \bibinfo{author}{\bibfnamefont{H.}~\bibnamefont{Neven}}, \bibnamefont{and}
  \bibinfo{author}{\bibfnamefont{J.~R.} \bibnamefont{McClean}},
  \bibinfo{journal}{Nature Communications} \textbf{\bibinfo{volume}{12}},
  \bibinfo{pages}{2631} (\bibinfo{year}{2021}{\natexlab{b}}), ISSN
  \bibinfo{issn}{2041-1723},
  \urlprefix\url{https://doi.org/10.1038/s41467-021-22539-9}.

\bibitem[{\citenamefont{Schuld}(2021)}]{schuld2021quantum}
\bibinfo{author}{\bibfnamefont{M.}~\bibnamefont{Schuld}}
  (\bibinfo{year}{2021}), \eprint{2101.11020},
  \urlprefix\url{http://arxiv.org/abs/2101.11020}.

\bibitem[{\citenamefont{Mengoni and Di~Pierro}(2019)}]{mengoni2019kernel}
\bibinfo{author}{\bibfnamefont{R.}~\bibnamefont{Mengoni}} \bibnamefont{and}
  \bibinfo{author}{\bibfnamefont{A.}~\bibnamefont{Di~Pierro}},
  \bibinfo{journal}{Quantum Machine Intelligence} \textbf{\bibinfo{volume}{1}},
  \bibinfo{pages}{65} (\bibinfo{year}{2019}).

\bibitem[{\citenamefont{Li et~al.}(2015)\citenamefont{Li, Liu, Xu, and
  Du}}]{li2015experimental}
\bibinfo{author}{\bibfnamefont{Z.}~\bibnamefont{Li}},
  \bibinfo{author}{\bibfnamefont{X.}~\bibnamefont{Liu}},
  \bibinfo{author}{\bibfnamefont{N.}~\bibnamefont{Xu}}, \bibnamefont{and}
  \bibinfo{author}{\bibfnamefont{J.}~\bibnamefont{Du}},
  \bibinfo{journal}{Physical review letters} \textbf{\bibinfo{volume}{114}},
  \bibinfo{pages}{140504} (\bibinfo{year}{2015}).

\bibitem[{\citenamefont{patent application for the method described in this
  manuscript has been submitted~by Pasqal.}()}]{qk_patent}
\bibinfo{author}{\bibfnamefont{A.}~\bibnamefont{patent application for the
  method described in this manuscript has been submitted~by Pasqal.}}

\bibitem[{\citenamefont{Wang}(2005)}]{wang2005support}
\bibinfo{author}{\bibfnamefont{L.}~\bibnamefont{Wang}},
  \emph{\bibinfo{title}{Support vector machines: theory and applications}},
  vol. \bibinfo{volume}{177} (\bibinfo{publisher}{Springer Science \& Business
  Media}, \bibinfo{year}{2005}).

\bibitem[{\citenamefont{Mehrkanoon et~al.}(2012)\citenamefont{Mehrkanoon,
  Falck, and Suykens}}]{mehrkanoon2012approximate}
\bibinfo{author}{\bibfnamefont{S.}~\bibnamefont{Mehrkanoon}},
  \bibinfo{author}{\bibfnamefont{T.}~\bibnamefont{Falck}}, \bibnamefont{and}
  \bibinfo{author}{\bibfnamefont{J.~A.} \bibnamefont{Suykens}},
  \bibinfo{journal}{IEEE transactions on neural networks and learning systems}
  \textbf{\bibinfo{volume}{23}}, \bibinfo{pages}{1356} (\bibinfo{year}{2012}).

\bibitem[{\citenamefont{Mehrkanoon and Suykens}(2015)}]{mehrkanoon2015learning}
\bibinfo{author}{\bibfnamefont{S.}~\bibnamefont{Mehrkanoon}} \bibnamefont{and}
  \bibinfo{author}{\bibfnamefont{J.~A.} \bibnamefont{Suykens}},
  \bibinfo{journal}{Neurocomputing} \textbf{\bibinfo{volume}{159}},
  \bibinfo{pages}{105} (\bibinfo{year}{2015}).

\bibitem[{\citenamefont{Lu et~al.}(2020)\citenamefont{Lu, Yin, Li, Sun, Yang,
  and Hou}}]{lu2020solving}
\bibinfo{author}{\bibfnamefont{Y.}~\bibnamefont{Lu}},
  \bibinfo{author}{\bibfnamefont{Q.}~\bibnamefont{Yin}},
  \bibinfo{author}{\bibfnamefont{H.}~\bibnamefont{Li}},
  \bibinfo{author}{\bibfnamefont{H.}~\bibnamefont{Sun}},
  \bibinfo{author}{\bibfnamefont{Y.}~\bibnamefont{Yang}}, \bibnamefont{and}
  \bibinfo{author}{\bibfnamefont{M.}~\bibnamefont{Hou}},
  \bibinfo{journal}{Journal of Industrial \& Management Optimization}
  \textbf{\bibinfo{volume}{16}}, \bibinfo{pages}{1481} (\bibinfo{year}{2020}).

\bibitem[{\citenamefont{Thompson and Stewart}(2002)}]{Thompson_Stewart_2002}
\bibinfo{author}{\bibfnamefont{J.~M.~T.} \bibnamefont{Thompson}}
  \bibnamefont{and} \bibinfo{author}{\bibfnamefont{H.~B.}
  \bibnamefont{Stewart}}, \emph{\bibinfo{title}{Nonlinear dynamics and chaos}}
  (\bibinfo{publisher}{Wiley}, \bibinfo{year}{2002}), \bibinfo{edition}{2nd}
  ed.

\bibitem[{\citenamefont{Mercer}(1909)}]{mercer1909xvi}
\bibinfo{author}{\bibfnamefont{J.}~\bibnamefont{Mercer}},
  \bibinfo{journal}{Philosophical transactions of the royal society of London.
  Series A, containing papers of a mathematical or physical character}
  \textbf{\bibinfo{volume}{209}}, \bibinfo{pages}{415} (\bibinfo{year}{1909}).

\bibitem[{\citenamefont{Boyd et~al.}(2004)\citenamefont{Boyd, Boyd, and
  Vandenberghe}}]{boyd2004convex}
\bibinfo{author}{\bibfnamefont{S.}~\bibnamefont{Boyd}},
  \bibinfo{author}{\bibfnamefont{S.~P.} \bibnamefont{Boyd}}, \bibnamefont{and}
  \bibinfo{author}{\bibfnamefont{L.}~\bibnamefont{Vandenberghe}},
  \emph{\bibinfo{title}{Convex optimization}} (\bibinfo{publisher}{Cambridge
  university press}, \bibinfo{year}{2004}).

\bibitem[{\citenamefont{Kuhn and Tucker}(2014)}]{kuhn2014nonlinear}
\bibinfo{author}{\bibfnamefont{H.~W.} \bibnamefont{Kuhn}} \bibnamefont{and}
  \bibinfo{author}{\bibfnamefont{A.~W.} \bibnamefont{Tucker}}, in
  \emph{\bibinfo{booktitle}{Traces and emergence of nonlinear programming}}
  (\bibinfo{publisher}{Springer}, \bibinfo{year}{2014}), pp.
  \bibinfo{pages}{247--258}.

\bibitem[{\citenamefont{P{\'{e}}rez-Salinas
  et~al.}(2020)\citenamefont{P{\'{e}}rez-Salinas, Cervera-Lierta, Gil-Fuster,
  and Latorre}}]{PerezSalinas2020datareuploading}
\bibinfo{author}{\bibfnamefont{A.}~\bibnamefont{P{\'{e}}rez-Salinas}},
  \bibinfo{author}{\bibfnamefont{A.}~\bibnamefont{Cervera-Lierta}},
  \bibinfo{author}{\bibfnamefont{E.}~\bibnamefont{Gil-Fuster}},
  \bibnamefont{and} \bibinfo{author}{\bibfnamefont{J.~I.}
  \bibnamefont{Latorre}}, \bibinfo{journal}{{Quantum}}
  \textbf{\bibinfo{volume}{4}}, \bibinfo{pages}{226} (\bibinfo{year}{2020}),
  ISSN \bibinfo{issn}{2521-327X},
  \urlprefix\url{https://doi.org/10.22331/q-2020-02-06-226}.

\bibitem[{\citenamefont{Caro et~al.}(2021)\citenamefont{Caro, Gil-Fuster,
  Meyer, Eisert, and Sweke}}]{Caro2021encodingdependent}
\bibinfo{author}{\bibfnamefont{M.~C.} \bibnamefont{Caro}},
  \bibinfo{author}{\bibfnamefont{E.}~\bibnamefont{Gil-Fuster}},
  \bibinfo{author}{\bibfnamefont{J.~J.} \bibnamefont{Meyer}},
  \bibinfo{author}{\bibfnamefont{J.}~\bibnamefont{Eisert}}, \bibnamefont{and}
  \bibinfo{author}{\bibfnamefont{R.}~\bibnamefont{Sweke}},
  \bibinfo{journal}{{Quantum}} \textbf{\bibinfo{volume}{5}},
  \bibinfo{pages}{582} (\bibinfo{year}{2021}), ISSN \bibinfo{issn}{2521-327X},
  \urlprefix\url{https://doi.org/10.22331/q-2021-11-17-582}.

\bibitem[{\citenamefont{Buhrman et~al.}(2001)\citenamefont{Buhrman, Cleve,
  Watrous, and De~Wolf}}]{buhrman2001quantum}
\bibinfo{author}{\bibfnamefont{H.}~\bibnamefont{Buhrman}},
  \bibinfo{author}{\bibfnamefont{R.}~\bibnamefont{Cleve}},
  \bibinfo{author}{\bibfnamefont{J.}~\bibnamefont{Watrous}}, \bibnamefont{and}
  \bibinfo{author}{\bibfnamefont{R.}~\bibnamefont{De~Wolf}},
  \bibinfo{journal}{Physical Review Letters} \textbf{\bibinfo{volume}{87}},
  \bibinfo{pages}{167902} (\bibinfo{year}{2001}).

\bibitem[{\citenamefont{Higgott et~al.}(2019)\citenamefont{Higgott, Wang, and
  Brierley}}]{Higgott2019variationalquantum}
\bibinfo{author}{\bibfnamefont{O.}~\bibnamefont{Higgott}},
  \bibinfo{author}{\bibfnamefont{D.}~\bibnamefont{Wang}}, \bibnamefont{and}
  \bibinfo{author}{\bibfnamefont{S.}~\bibnamefont{Brierley}},
  \bibinfo{journal}{{Quantum}} \textbf{\bibinfo{volume}{3}},
  \bibinfo{pages}{156} (\bibinfo{year}{2019}), ISSN \bibinfo{issn}{2521-327X},
  \urlprefix\url{https://doi.org/10.22331/q-2019-07-01-156}.

\bibitem[{\citenamefont{Mitarai and Fujii}(2019)}]{Mitarai2019Htest}
\bibinfo{author}{\bibfnamefont{K.}~\bibnamefont{Mitarai}} \bibnamefont{and}
  \bibinfo{author}{\bibfnamefont{K.}~\bibnamefont{Fujii}},
  \bibinfo{journal}{Phys. Rev. Research} \textbf{\bibinfo{volume}{1}},
  \bibinfo{pages}{013006} (\bibinfo{year}{2019}),
  \urlprefix\url{https://link.aps.org/doi/10.1103/PhysRevResearch.1.013006}.

\bibitem[{\citenamefont{Kyriienko and
  Elfving}(2021)}]{kyriienko2021generalized}
\bibinfo{author}{\bibfnamefont{O.}~\bibnamefont{Kyriienko}} \bibnamefont{and}
  \bibinfo{author}{\bibfnamefont{V.~E.} \bibnamefont{Elfving}},
  \bibinfo{journal}{Physical Review A} \textbf{\bibinfo{volume}{104}},
  \bibinfo{pages}{052417} (\bibinfo{year}{2021}).

\bibitem[{\citenamefont{Wierichs et~al.}(2021)\citenamefont{Wierichs, Izaac,
  Wang, and Lin}}]{wierichs2021general}
\bibinfo{author}{\bibfnamefont{D.}~\bibnamefont{Wierichs}},
  \bibinfo{author}{\bibfnamefont{J.}~\bibnamefont{Izaac}},
  \bibinfo{author}{\bibfnamefont{C.}~\bibnamefont{Wang}}, \bibnamefont{and}
  \bibinfo{author}{\bibfnamefont{C.~Y.-Y.} \bibnamefont{Lin}},
  \bibinfo{journal}{arXiv preprint arXiv:2107.12390}  (\bibinfo{year}{2021}).

\bibitem[{\citenamefont{Izmaylov et~al.}(2021)\citenamefont{Izmaylov, Lang, and
  Yen}}]{Izmaylov2021generalized}
\bibinfo{author}{\bibfnamefont{A.~F.} \bibnamefont{Izmaylov}},
  \bibinfo{author}{\bibfnamefont{R.~A.} \bibnamefont{Lang}}, \bibnamefont{and}
  \bibinfo{author}{\bibfnamefont{T.-C.} \bibnamefont{Yen}},
  \bibinfo{journal}{Phys. Rev. A} \textbf{\bibinfo{volume}{104}},
  \bibinfo{pages}{062443} (\bibinfo{year}{2021}),
  \urlprefix\url{https://link.aps.org/doi/10.1103/PhysRevA.104.062443}.

\bibitem[{\citenamefont{Vidal and Theis}(2018)}]{Vidal2018}
\bibinfo{author}{\bibfnamefont{J.~G.} \bibnamefont{Vidal}} \bibnamefont{and}
  \bibinfo{author}{\bibfnamefont{D.~O.} \bibnamefont{Theis}}
  (\bibinfo{year}{2018}), \eprint{1812.06323},
  \urlprefix\url{https://arxiv.org/abs/1812.06323}.

\bibitem[{\citenamefont{Theis}(2021)}]{Theis2021}
\bibinfo{author}{\bibfnamefont{D.~O.} \bibnamefont{Theis}}
  (\bibinfo{year}{2021}), \eprint{2112.14669},
  \urlprefix\url{https://arxiv.org/abs/2112.14669}.

\bibitem[{\citenamefont{Savary and Balents}(2016)}]{Savary2016}
\bibinfo{author}{\bibfnamefont{L.}~\bibnamefont{Savary}} \bibnamefont{and}
  \bibinfo{author}{\bibfnamefont{L.}~\bibnamefont{Balents}},
  \bibinfo{journal}{Reports on Progress in Physics}
  \textbf{\bibinfo{volume}{80}}, \bibinfo{pages}{016502}
  (\bibinfo{year}{2016}),
  \urlprefix\url{https://doi.org/10.1088/0034-4885/80/1/016502}.

\bibitem[{\citenamefont{Hermanns et~al.}(2018)\citenamefont{Hermanns, Kimchi,
  and Knolle}}]{Hermanns2018}
\bibinfo{author}{\bibfnamefont{M.}~\bibnamefont{Hermanns}},
  \bibinfo{author}{\bibfnamefont{I.}~\bibnamefont{Kimchi}}, \bibnamefont{and}
  \bibinfo{author}{\bibfnamefont{J.}~\bibnamefont{Knolle}},
  \bibinfo{journal}{Annual Review of Condensed Matter Physics}
  \textbf{\bibinfo{volume}{9}}, \bibinfo{pages}{17} (\bibinfo{year}{2018}),
  \eprint{https://doi.org/10.1146/annurev-conmatphys-033117-053934},
  \urlprefix\url{https://doi.org/10.1146/annurev-conmatphys-033117-053934}.

\bibitem[{\citenamefont{Bespalova and
  Kyriienko}(2021{\natexlab{a}})}]{bespalova2021quantum}
\bibinfo{author}{\bibfnamefont{T.~A.} \bibnamefont{Bespalova}}
  \bibnamefont{and}
  \bibinfo{author}{\bibfnamefont{O.}~\bibnamefont{Kyriienko}},
  \bibinfo{journal}{arXiv preprint arXiv:2109.13883}
  (\bibinfo{year}{2021}{\natexlab{a}}).

\bibitem[{\citenamefont{Huang et~al.}(2021{\natexlab{c}})\citenamefont{Huang,
  Broughton, Cotler, Chen, Li, Mohseni, Neven, Babbush, Kueng, Preskill
  et~al.}}]{Huang2021b}
\bibinfo{author}{\bibfnamefont{H.-Y.} \bibnamefont{Huang}},
  \bibinfo{author}{\bibfnamefont{M.}~\bibnamefont{Broughton}},
  \bibinfo{author}{\bibfnamefont{J.}~\bibnamefont{Cotler}},
  \bibinfo{author}{\bibfnamefont{S.}~\bibnamefont{Chen}},
  \bibinfo{author}{\bibfnamefont{J.}~\bibnamefont{Li}},
  \bibinfo{author}{\bibfnamefont{M.}~\bibnamefont{Mohseni}},
  \bibinfo{author}{\bibfnamefont{H.}~\bibnamefont{Neven}},
  \bibinfo{author}{\bibfnamefont{R.}~\bibnamefont{Babbush}},
  \bibinfo{author}{\bibfnamefont{R.}~\bibnamefont{Kueng}},
  \bibinfo{author}{\bibfnamefont{J.}~\bibnamefont{Preskill}},
  \bibnamefont{et~al.} (\bibinfo{year}{2021}{\natexlab{c}}),
  \eprint{2112.00778}, \urlprefix\url{https://arxiv.org/abs/2112.00778}.

\bibitem[{\citenamefont{Luo et~al.}(2019)\citenamefont{Luo, Liu, Zhang, and
  Wang}}]{YaoFramework2019}
\bibinfo{author}{\bibfnamefont{X.-Z.} \bibnamefont{Luo}},
  \bibinfo{author}{\bibfnamefont{J.-G.} \bibnamefont{Liu}},
  \bibinfo{author}{\bibfnamefont{P.}~\bibnamefont{Zhang}}, \bibnamefont{and}
  \bibinfo{author}{\bibfnamefont{L.}~\bibnamefont{Wang}},
  \bibinfo{journal}{arXiv preprint arXiv:1912.10877}  (\bibinfo{year}{2019}).

\bibitem[{\citenamefont{Larocca
  et~al.}(2021{\natexlab{a}})\citenamefont{Larocca, Ju,
  Garc{\'{i}}a-Mart{\'{i}}n, Coles, and
  Cerezo}}]{Larocca2021overparametrization}
\bibinfo{author}{\bibfnamefont{M.}~\bibnamefont{Larocca}},
  \bibinfo{author}{\bibfnamefont{N.}~\bibnamefont{Ju}},
  \bibinfo{author}{\bibfnamefont{D.}~\bibnamefont{Garc{\'{i}}a-Mart{\'{i}}n}},
  \bibinfo{author}{\bibfnamefont{P.~J.} \bibnamefont{Coles}}, \bibnamefont{and}
  \bibinfo{author}{\bibfnamefont{M.}~\bibnamefont{Cerezo}}
  (\bibinfo{year}{2021}{\natexlab{a}}), \eprint{2109.11676},
  \urlprefix\url{http://arxiv.org/abs/2109.11676}.

\bibitem[{\citenamefont{Larocca
  et~al.}(2021{\natexlab{b}})\citenamefont{Larocca, Czarnik, Sharma,
  Muraleedharan, Coles, and Cerezo}}]{Larocca2021optimal}
\bibinfo{author}{\bibfnamefont{M.}~\bibnamefont{Larocca}},
  \bibinfo{author}{\bibfnamefont{P.}~\bibnamefont{Czarnik}},
  \bibinfo{author}{\bibfnamefont{K.}~\bibnamefont{Sharma}},
  \bibinfo{author}{\bibfnamefont{G.}~\bibnamefont{Muraleedharan}},
  \bibinfo{author}{\bibfnamefont{P.~J.} \bibnamefont{Coles}}, \bibnamefont{and}
  \bibinfo{author}{\bibfnamefont{M.}~\bibnamefont{Cerezo}}
  (\bibinfo{year}{2021}{\natexlab{b}}), \eprint{2105.14377},
  \urlprefix\url{https://arxiv.org/abs/2105.14377}.

\bibitem[{\citenamefont{Bespalova and
  Kyriienko}(2021{\natexlab{b}})}]{Bespalova2021}
\bibinfo{author}{\bibfnamefont{T.~A.} \bibnamefont{Bespalova}}
  \bibnamefont{and}
  \bibinfo{author}{\bibfnamefont{O.}~\bibnamefont{Kyriienko}},
  \bibinfo{journal}{PRX Quantum} \textbf{\bibinfo{volume}{2}},
  \bibinfo{pages}{030318} (\bibinfo{year}{2021}{\natexlab{b}}),
  \urlprefix\url{https://link.aps.org/doi/10.1103/PRXQuantum.2.030318}.

\end{thebibliography}

\appendix*
\section{Formulating SVR Problem}
Here, we show how to formulate an SVR problem, providing more details for the specifical example. We consider the case $\mathrm{DE}(x, f, df/dx) = df/dx - g(x, f) = 0$ with initial condition $f(x_0) = f_0$. We use the model formulated as $f(x) = \mathbf{w}^\dag \bm{\varphi}(x) + b$.

As a first step, the problem needs to be written in primal SVR model form,
\begin{align}
    &\mathrm{min}_{e, \xi, w, b} \mathbf{w}^\dag\mathbf{w} + \gamma \mathbf{e}^T \mathbf{e} + \gamma \bm{\xi}^T \bm{\xi},\\
    &\mathrm{subject~to}~ \mathbf{w}^T \bm{\varphi}'(x_i) - g(x_i, y_i) = e_i ~ i=1:N,\\
    &\mathbf{w}^T \mathbf{\phi}(x_0) + b = f_0,\\
    &y_i = \mathbf{w}^T\bm{\varphi}(x_i) + b + \xi_i ~ i =1:N .
\end{align}
Here, the minimization function is such that the magnitude of $\mathbf{w}$, $\mathbf{e}$ and $\bm{\xi}$ are minimized, with $\mathbf{w}$ being the set of fitting coefficients. $\mathbf{e}$ and $\bm{\xi}$ are the errors in the constraints. Minimizing this function one finds the smallest $\mathbf{w}$ that fulfils the constraints with smallest possible error. Finding the smallest possible $\mathbf{w}$ is a form of regularisation helping prevent overfitting. $\gamma$ is a tunable hyperparameter which dictates how much emphasis is placed on error reduction.

The constraints correspond to the differential equation at each point $x_i$, the initial condition and introduced dummy variables $y_i = f(x_i) + \xi_i$, respectively. The dummy variables are introduced to reflect the nonlinearity of the problem. 

The second step is to find the Lagrangian of the model. This corresponds to the minimization function minus each of the constraints, preceded by a variable coefficient,
\begin{align}
    \mathcal{L} =& \frac{1}{2}\mathbf{w}^T\mathbf{w} + \frac{\gamma}{2} \mathbf{e}^T \mathbf{e} + \frac{\gamma}{2} \bm{\xi}^T \xi \\
    &- \sum_{i=1}^N \alpha_i (\mathbf{w}^T \bm{\varphi}'(x_i) - g(x_i, y_i) - e_i) \\
    &- \beta (\mathbf{w}^T \mathbf{\varphi}(x_0) + b - f_0) \\
    &- \sum_{i=1}^N \eta_i (\mathbf{w}^T\bm{\varphi}(x_i) + b + \xi_i - y_i).
\end{align}
The introduced variables $\bm{\alpha}$, $\bm{\nu}$ and $\beta$ are referred to as dual variables.

The next step is to calculate the KKT conditions. These are found by equating to zero derivative of the Lagrangian with respect to each of its variables, both primal and dual, ($\mathbf{w}, b, \mathbf{e}, \bm{\xi}, \mathbf{y}, \bm{\alpha}, \beta, \bm{\eta}$). The derivatives read
\begin{align}
    \frac{\partial \mathcal{L}}{\partial \mathbf{w}} &= \mathbf{w} - \sum_i \left( \alpha_i \bm{\varphi}'(x_i) + \eta_i \bm{\varphi}(x_i) \right) - \beta \bm{\varphi}(x_0)  = 0, \\
    \frac{\partial \mathcal{L}}{\partial b} &= - \beta - \sum_i \eta_i = 0 ,\\
    \frac{\partial \mathcal{L}}{\partial e_i} &= \gamma e_i + \alpha_i = 0 ,\\
    \frac{\partial \mathcal{L}}{\partial \xi_i} &= \gamma \xi_i - \eta_i = 0 ,\\
    \frac{\partial \mathcal{L}}{\partial y_i} &= \alpha_i \frac{\partial g}{\partial y}(x_i, y_i) + \eta_i = 0 ,\\
    \frac{\partial \mathcal{L}}{\partial \alpha_i} &= - \left( \mathbf{w}^\dag \bm{\varphi}'(x_i) - g(x_i, y_i) - e_i \right) = 0 ,\\
    \frac{\partial \mathcal{L}}{\partial \beta} &=  - \left(\mathbf{w}^\dag \bm{\varphi}(x_0) + b - f_0 \right) = 0 , \\
    \frac{\partial \mathcal{L}}{\partial \nu_i} &=  - \left(y_i - \mathbf{w}^\dag \bm{\varphi}(x_i) - b - \xi_i \right) = 0.
\end{align}
These are a set of $6|\mathbf{x}|+2$ equations which necessarily need to be satisfied for optimality. 

These conditions are now used to eliminate a subset of the primal variables $\mathbf{w}, \mathbf{e}, \bm{\xi}$ leaving $3|\mathbf{x}|+2$ equations:
\begin{align}
    \left( \sum_j [\alpha_j \bm{\varphi}'(x_j) + \nu_j \bm{\varphi}(x_j)]+ \beta \bm{\varphi}(x_0)\right)^\dag \bm{\varphi}'(x_i) \\ - g(x_i, y_i) + \alpha_i/\gamma = 0 , \\
    \left( \sum_j [\alpha_j \bm{\varphi}'(x_j) + \nu_j \bm{\varphi}(x_j)]+ \beta \bm{\varphi}(x_0)\right)^\dag \bm{\varphi}(x_0) \\ + b - f_0 = 0 , \\
    -\left( \sum_j [\alpha_j \bm{\varphi}'(x_j) + \nu_j \bm{\varphi}(x_j)]+ \beta \bm{\varphi}(x_0)\right)^\dag \bm{\varphi}(x_i) \\
    + y_i -b - \eta_i/\gamma = 0  ,\\
    \sum_i \eta_i + \beta = 0  ,\\
    \alpha_i \frac{\partial g}{\partial y}(x_i, y_i) + \eta_i = 0.
\end{align}
For these equations we then expand out the brackets and use the kernel trick, introducing the kernel function $\kappa$ as $\kappa(x, y) = \bm{\varphi}^\dag(x)\bm{\varphi}(y)$ and corresponding derivatives. We remember that this is a consequence of Mercers theorem, given that $\bm{\varphi}^\dag(x)\bm{\varphi}(y)$ is a kernel for any $\bm{\varphi}$. Now we are able to write the resulting equations in matrix form as
\begin{align}
    \left[\begin{array}{c|c|c|c|c} \tilde{\Omega}_1^1 & \Omega_0^{1} & \mathbf{h}_0^1 & \mathbf{0} & \hat{0} \\
    \hline
    \Omega_1^{0} & \tilde{\Omega_0^0} & \mathbf{h}_0^0 & \mathbf{1} & -I \\
    \hline
    {\mathbf{h}^T}_1^0 & {\mathbf{h}^T}_0^{0} & \tilde{h} & 1 & \mathbf{0}^T \\
    \hline
    \mathbf{0}^T & \mathbf{1}^T & 1 & 0 & \mathbf{0}^T \\
    \hline
    \hat{D} & I & \mathbf{0} & \mathbf{0} & \mathbf{0}
    \end{array}\right] \left[ \begin{array}{c} 
    \bm{\alpha} \\ \hline \bm{\eta} \\ \hline \beta \\ \hline b \\ \hline \mathbf{y}
    \end{array} \right]= 
    \left[ \begin{array}{c} 
    \mathbf{\tilde{g}} \\ \hline \mathbf{0} \\ \hline f_0 \\ \hline 0 \\ \hline \hat{0}
    \end{array} \right],
\end{align}
where the notation is as follows
\begin{align}
    [\Omega^m_n]_{i,j} &= \nabla^m_n \kappa(x_j, x_i) , \\
    \tilde{\Omega}^m_n &= \Omega^m_n + \hat{I}/\gamma , \\
    [\mathbf{h}^m_n]_i &= \nabla^m_n \kappa(x_0, x_i),  \\
    \tilde{h} &= \kappa(x_0, x_0),  \\
    \hat{D} &= \mathrm{diag}\left( \left\{ \frac{\partial g}{\partial f}(x_i, y_i) \right \}_i \right),\\
    [\mathbf{\tilde{g}}]_i &= g(x_i, y_i) .
\end{align}
We now have a set of nonlinear equations that can be solved for a set of variable, representing solution to the original stated problem. These equations are written in terms of $\kappa$, and not $\bm{\varphi}$. Also note that these equations are true for any valid kernel function, and we can choose our kernel function freely. We need not know what the corresponding $\bm{\varphi}$ are, we simply know from Mercers theorem that such functions exist. Therefore the formulation of these equations (in particular the use of the kernel trick to introduce the kernel) is valid.

The remaining step is to write $f(x) = \mathbf{w}^\dag \bm{\varphi}(x) + b$ in a form that is instead dependent on the variables solved for. We find it to be %
\begin{align}
    f(x) = \sum_{i=1}^{|\mathbf{x}|} \alpha_i \nabla_1^0 \kappa(x_i, x) + \sum_{i=1}^{|\mathbf{x}|} \eta_i \kappa(x_i, x) + \beta \kappa(x_0, x) + b,
\end{align}
by using the $\mathbf{w}$ KKT condition and then the kernel trick. We have now formulated an SVR method for the form of problem considered.
\end{document}